\documentclass[authoryear,preprint,review,12pt]{elsarticle}

\usepackage{amsmath}
\usepackage{amsthm}
\usepackage{amssymb}
\usepackage{graphicx}
\usepackage{textcomp}
\usepackage{xcolor}
\usepackage{amsfonts}
\usepackage{bbm}
\usepackage{tabularx}
\usepackage{diagbox}
\usepackage{booktabs}
\usepackage{threeparttable}
\usepackage{subfigure}
\usepackage{caption}
\usepackage{multirow}
\usepackage{makecell}
\usepackage{float}
\usepackage{rotating}

\theoremstyle{definition}

\newtheorem{remark}{Remark}

\newtheorem{assumption}{Assumption}

\newtheorem{prop}{Proposition}
\newtheorem{lem}{Lemma}
\newtheorem{cor}{Corollary}

\newcommand{\cs}[1]{\texttt{\symbol{`\\}#1}}

\usepackage{algorithm}
\usepackage{algpseudocode}
\usepackage{hyperref}
\let\oldReturn\Return
\renewcommand{\Return}{\State\oldReturn}
\makeatletter
\newcommand{\multiline}[1]{%
  \begin{tabularx}{\dimexpr\linewidth-\ALG@thistlm}[t]{@{}X@{}}
    #1
  \end{tabularx}
}
\makeatother

\journal{Neural Networks}

\begin{document}

\begin{frontmatter}



\title{Communication-Efficient Diffusion Strategy for Performance Improvement of Federated Learning with Non-IID Data}


\author[inst1]{Seyoung Ahn}
\author[inst1]{Soohyeong Kim}
\author[inst1]{Yongseok Kwon}
\author[inst1]{Jiseung Youn}
\author[inst3]{Joohan Park}
\author[inst2]{Sunghyun Cho}

\affiliation[inst1]{organization={Department of Computer Science and Engineering, Major in Bio Artificial Intelligence, Hanyang University},
            postcode={15588},
            country={South Korea}}

\affiliation[inst2]{organization={Department of Computer Science and Engineering, Hanyang University ERICA},
            postcode={15588},
            country={South Korea}}

\affiliation[inst3]{organization={Spatial Wireless Networking Research Section, Electronics and Telecommunications Research Institute (ETRI)},
            postcode={34129},
            country={South Korea}}
            
\begin{abstract}
In 6G mobile communication systems, various AI-based network functions and applications have been standardized. Federated learning (FL) is adopted as the core learning architecture for 6G systems to avoid privacy leakage from mobile user data. However, in FL, users with non-independent and identically distributed (non-IID) datasets can deteriorate the performance of the global model because the convergence direction of the gradient for each dataset is different, thereby inducing a weight divergence problem. To address this problem, we propose a novel diffusion strategy for machine learning (ML) models (FedDif) to maximize the performance of the global model with non-IID data. FedDif enables the local model to learn different distributions before parameter aggregation by passing the local models through users via device-to-device communication. Furthermore, we theoretically demonstrate that FedDif can circumvent the weight-divergence problem. Based on this theory, we propose a communication-efficient diffusion strategy for ML models that can determine the trade-off between learning performance and communication cost using auction theory. The experimental results show that FedDif improves the top-1 test accuracy by up to 34.89\% and reduces communication costs by 14.6\% to a maximum of 63.49\%.
\end{abstract}

\begin{graphicalabstract}
\includegraphics[width=\textwidth]{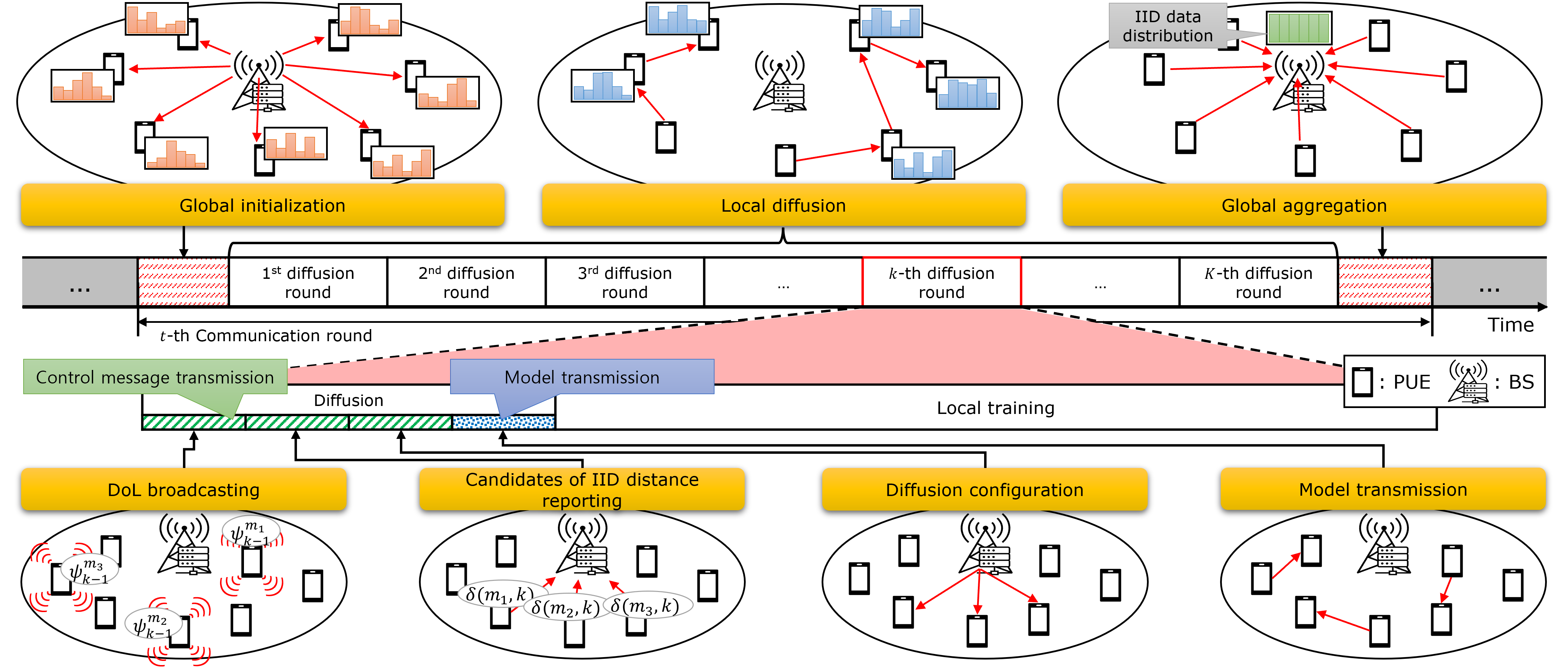}
\end{graphicalabstract}

\begin{highlights}
\item We introduce a novel diffusion mechanism for machine learning models called \emph{FedDif}, which aims to reduce weight divergence caused by non-IID data.
\item We design the diffusion strategy based on auction theory to balance the enhancement of learning performance with the reduction of communication costs.
\item We provide a theoretical analysis of FedDif demonstrating that the diffusion mechanism can mitigate weight divergence.
\item We further demonstrate empirically that FedDif improves the performance of the global model and communication efficiency through simulations.
\end{highlights}


\end{frontmatter}


\section{Introduction}\label{sec:intro}
As interest in user data privacy has increased, a novel distributed learning architecture has been proposed for users to participate in the training process. Federated learning (FL) is a representative scheme used for distributed learning without privacy leakage~\citep{ref:fedavg}. It preserves the privacy of user data through two procedures: \emph{local training} and \emph{global aggregation}. A central server distributes the global model to the users, and the users train the local model using their private data (\emph{local training}). Furthermore, the central server collects the local models from the users and updates the global model (\emph{global aggregation}). Global aggregation is an essential part of FL for capturing the advantages of maintaining learning performance and preserving user data privacy.\par
Although FL has many advantages in training neural networks, it faces a crucial challenge for uncharted user data. While the central server can directly curate a training dataset in centralized learning, it cannot manage the data for each user. Specifically, non-independent and identically distributed (non-IID) data can cause severe performance degradation~\citep{ref:noniid2}. If users train the local model with their non-IID data, they can be trained on a biased optimization trajectory owing to the heterogeneous data distribution. The direction of the biased trajectory may not head toward the global optimum, and neither can the optimization trajectory of the aggregated global model. The weight divergence of the global model indicates this problem~\citep{ref:noniid1, ref:noniid3}.\par
Preventing the performance degradation of the global model depends on reducing the impact of non-IID data by aligning the biased optimization trajectory with the global optimum. As one of the methods, data augmentation to supplement unbalanced datasets using the public dataset or generative model can prevent the bias of the optimization trajectory~\citep{ref:noniid1, ref:noniid3, ref:feddistil}. Because data-augmentation schemes are typically based on user data, data privacy can be compromised. Regularization is another typical approach for addressing overfitting, and many state-of-the-art (SOTA) methods utilize it to mitigate the impact of non-IID data on FL~\citep{ref:SCAFFOLD, ref:fedprox, ref:FedDyn, ref:FedDC}. Combining the advantages of data augmentation and regularization without compromising data privacy may achieve faster convergence speeds and even higher performance improvements.\par
We propose a novel diffusion strategy for machine learning (ML) models, called FedDif to improve the performance of a global model with non-IID user data and preserve data privacy. First, we discover some insights for designing FedDif by asking, \emph{``Is it possible to achieve the performance equivalent to training on an IID dataset by training various non-IID data puzzles on local models before global aggregation?''} We rethink the general concept of FL in which the user trains the ML model based on the fact that the model learns different users’ data. In FedDif, each user is considered as non-IID batch data, and local models learn various non-IID data by passing them through multiple users before global aggregation. We propose a diffusion mechanism in which users relay models for training with more diverse data. Moreover, we propose a novel concept, the IID distance, to measure how the local model learns from various non-IID datasets to approach the IID distribution. Through the diffusion mechanism, the models seek other trainers to place non-IID data pieces for the IID data puzzle by minimizing the IID distance. Consequently, we can achieve an effect similar to that of a model that trains IID data via several diffusion iterations.\par
Although the diffusion mechanism can mitigate the effects of non-IID data, excessive diffusion can substantially increase the total training time and deteriorate the performance of communication systems. In other words, there is a trade off between improving learning performance and reducing communication costs. Immoderate diffusion can deteriorate network performance because users can overoccupy the bandwidth required to send their model. Conversely, passive diffusion requires more time-domain resources to achieve the targeted performance. Therefore, an efficient scheduling method should be designed for communication-efficient diffusion by considering these trade-offs. We first construct an optimization problem between IID distance and communication cost to find the optimal scheduling policy. Then, we provide a theoretical analysis of FedDif by demonstrating that FedDif can mitigate weight divergence. Our analysis provides a guideline for optimization, in which the scheduling policy should assign the next users who can minimize the IID distance of the models. Finally, we propose a diffusion strategy to find a feasible solution to the optimization problem based on auction theory.\par
Our contributions can be summarized as follows:
\begin{itemize}
    \item We introduce a novel diffusion mechanism for machine learning models called \emph{FedDif}, which aims to reduce weight divergence caused by non-IID data. In this mechanism, local models accumulate the personalized data distributions from different users, achieving a similar effect to training on IID data.
    \item We design the diffusion strategy based on auction theory to balance the enhancement of learning performance with the reduction of communication costs. We formulate an optimization problem to find the trade-off, and the auction provides a feasible solution based on the proposed winner selection algorithm.
    \item We provide a theoretical analysis of FedDif demonstrating that the diffusion mechanism can mitigate weight divergence. Two propositions show that the IID distance is closely related to weight divergence, and the diffusion mechanism can minimize the IID distance of the model after sufficient diffusion.
    \item We further demonstrate empirically that FedDif improves the performance of the global model and communication efficiency through simulations. FedDif increases the test accuracy of the global model while minimizing communication costs compared to other SOTA methods. Additionally, we provide our implementations of FedDif in an open-source code repository\footnote{The official implementations of FedDif are available at \href{https://github.com/seyoungahn/FedDif}{https://github.com/seyoungahn/FedDif}}.
\end{itemize}

The remainder of this paper is organized as follows. Section \ref{section2} introduces the related work. Section \ref{section3} provides the system model and problem formulation. Section \ref{section5} proposes FedDif based on auction theory. Sections \ref{section4} and \ref{section6} present the theoretical and experimental analysis of FedDif, respectively. Finally, we conclude this paper in Section \ref{section7}.

\section{Related works}\label{section2}
Recently, several SOTA methods have been proposed to address the weight-divergence problem for non-IID data by regularization~\citep{ref:SCAFFOLD, ref:fedprox, ref:FedDyn, ref:FedDC} and user cooperation~\citep{ref:AIoT, ref:selected1, ref:coopfl1, ref:coopfl-revision1, ref:FLT}. SCAFFOLD~\citep{ref:SCAFFOLD} corrects the gradient direction by penalizing the loss function with a gradient drift term. Using the gradient correction proposed by SCAFFOLD, FedDC~\citep{ref:FedDC} conflates insights from previous studies on regularization, and exhibits state-of-the-art performance.
\citet{ref:AIoT} proposed a novel operation called FedSwap, which randomly re-distributes local models to users to train on multiple non-IID datasets before aggregation. \citet{ref:selected1} proposed two-timescale hybrid federated learning (TT-HF), which aggregates the model within local clusters and cooperatively trains the model in each local cluster. \citet{ref:coopfl1} proposed consensus-based federated averaging (CFA), wihch trains the ML model through consensus-based in-network federated learning between different users via D2D communications. \citet{ref:FLT} proposed a user selection method using taskonomy, which groups users based on task-relatedness among them.\par
Several studies have been explored to provide advanced functionalities of FL such as communication efficiency and security. \citet{ref:coopfl-revision1} also investigated several methods of decentralized FL. The authors emphasized that exploring the trade-off between convergence speed and the cost of a large communication overhead is currently an open challenge. \citet{ref:selected2} proposed the model compression framework based on sparse ternary compression (STC) that meets the requirements of FL. By compressing the models with STC, they significantly reduced the communication costs. The works in \citep{ref:EPPDA, ref:sec_agg, ref:ParSGD} claim additional privacy leakage to the exchanged parameters in the global aggregation step, and performance degradation owing to user failure. These works propose fault-tolerant aggregation methods, such as partial synchronous SGD and secure aggregation schemes with secret sharing based on multi-party computation (MPC). These methods are compatible with our proposed FedDif, which enhances the robustness of FedDif and create a synergy effect with FedDif.\par
Although the above studies on FL with non-IID data have improved the performance of the global model and ensured advanced functionalities such as security and communication efficiency, it is still challenging to simultaneously address privacy leakage, performance degradation, and communication overhead. Therefore, it is necessary to combine the strengths of the previous studies to ensure data privacy, communication efficiency, and performance improvement simultaneously. We propose a convergence strategy for regularization and data augmentation through user cooperation to put the non-IID data puzzles without compromising user data privacy. We also ensure communication efficiency by optimizing the diffusion efficiency. We further clarify the reason for weight divergence through a theoretical analysis.\par

\begin{figure*}[t]
  \centering
  \includegraphics[width=\textwidth]{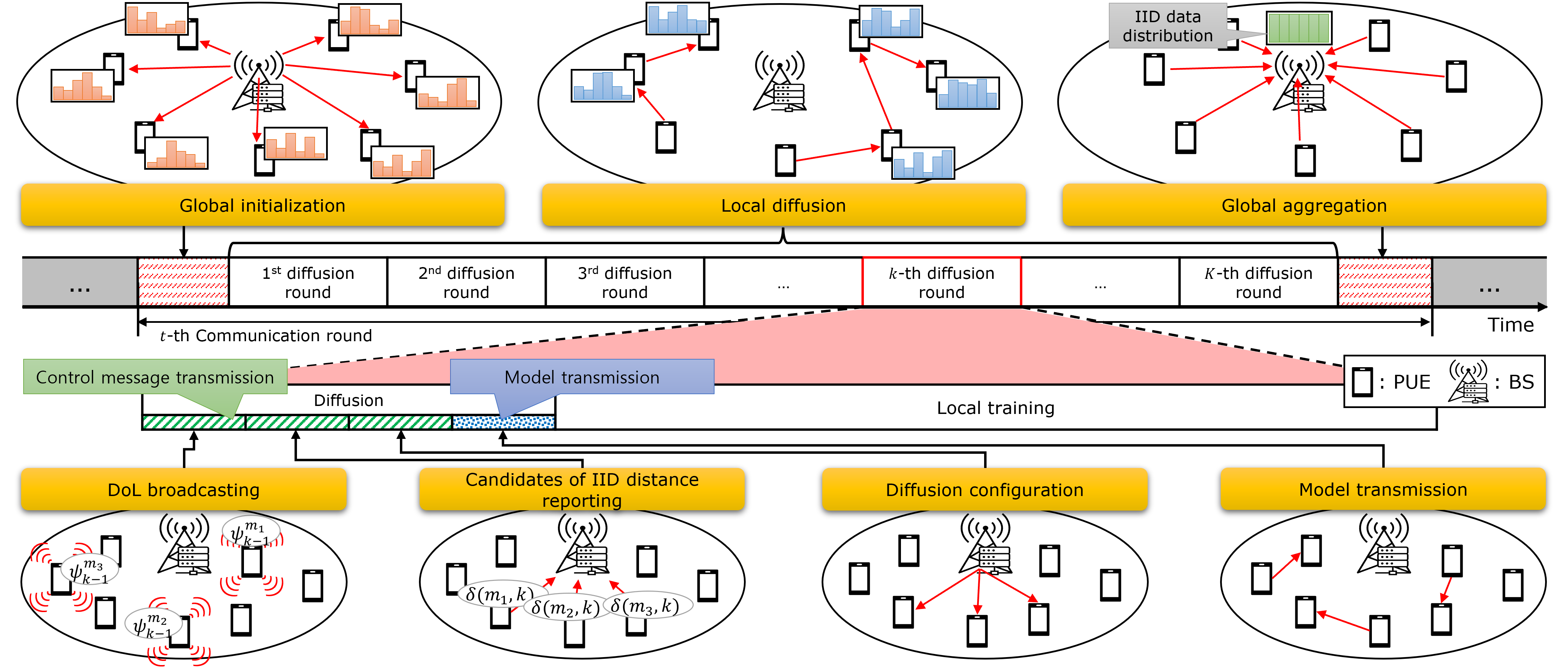}
    \caption{Overview of FedDif.}
  \label{fig:diffusion_concept}
\end{figure*}

\section{System model and problem formulation}\label{section3}
This section describes the system model FedDif operates (Section \ref{section3A}), and the representations of the data distribution (Section \ref{section3B}). Then, we formulate an optimization problem to determine the trade-off between learning performance and communication cost (Section \ref{section3C}). 

\subsection{System description}\label{section3A}
We consider typical FL frameworks for wireless networks consisting of a base station (BS) and user equipment (UE). Assuming that $T$ communication rounds are required to train the global model sufficiently, the BS first selects the available UE subset to prevent failure of training the global model owing to the outage of the UE
\footnote{Several methods can be used for selecting available UEs. The BS can either select UEs before initiating the FL process or identify and choose UEs with available communication and computing capabilities before each communication round. In this study, we adopt the latter approach.}.
We define a subset of $N_{P}$ available UEs as the participating UE (PUE) set $\mathcal{N}_{P}=\{1,...,N_{P}\}$. UEs that do not participate in FL comprise the cellular UE (CUE) set $\mathcal{N}_{C}=\{1,...,N_{C}\}$. The BS distributes the global model $\textbf{w}_{t-1}^{(g)}$ comprising the parameters of the global model in the $(t-1)$th communication round to PUEs. $\mathcal{M}$ is a set of distributed models called the local model, and $m \in \mathcal{M}$ is the local model with parameters $\textbf{w}_{t, 0}^{(m)}$ replicated by the global model. Local models are independent of PUEs in FedDif, and each PUE is considered a single batch of data.\par

Unlike in a typical FL, in FedDif, PUEs diffuse models to neighboring PUEs before global aggregation, as illustrated in Fig.~\ref{fig:diffusion_concept}. We define one diffusion iteration as a diffusion round. In the $k$-th diffusion round, each PUE $i$ trains the local model $\textbf{w}_{t, k}^{(m)}$ using its private dataset $\mathcal{D}_{i}$ of size $D_{i}$. The dataset of PUE $i$ is generated using different random variables $X_{i}$. The samples in each dataset are in the label space $\mathcal{C}$ whose cardinality is $C$. To define centralized learning with IID data, we consider a universal dataset comprising the dataset of all PUEs represented as $\mathcal{D}=\bigcup_{i \in \mathcal{N}_{p}} \mathcal{D}_{i}$. The total data size of the system is $D=\sum_{i \in \mathcal{N}_{P}}D_{i}$. The universal dataset is generated using a single random variable $X_{g}$. The probability density functions (PDFs) of the random variables $X_{1}, ..., X_{N_{P}}, X_{g}$ are denoted by $P(X_{1}), ..., P(X_{N_{P}}), P(X_{g})$. We consider the relationship between PDFs of each PUE as $P(X_{i}) \neq P(X_{j})$, $i, j \in \{1, ..., N_{P}\}$, $i \neq j$ because each PUE's dataset is non-IID.\par

We define the diffusion chain $\mathcal{P}_{K_{t}}^{(m)}$ as a set of PUEs that participate in training model $m$. Here, $K_{t}$ denotes the last diffusion round in $t$-th communication round. PUEs can be members of different diffusion chains throughout the entire diffusion round because they require certain PUE datasets. We assume that PUEs can only participate in training the model $m$ once to avoid overtraining of their data distribution and train one model in a diffusion round because of their limited computational capacities. The total size of data managed by PUEs in the diffusion chain $\mathcal{P}^{(m)}_{K_{t}}$ is represented by $D_{(\mathcal{P}_{K_{t}}^{(m)})}=\sum_{i \in \mathcal{P}_{K_{t}}^{(m)}}D_{i}$. Furthermore, we define the diffusion subchain $\mathcal{P}_{k}^{(m)} \subseteq \mathcal{P}_{K_{t}}^{(m)}$ as a set of PUEs participating in the training of model $m$ up to the $k$th diffusion round. The total size of data in the diffusion subchain $\mathcal{P}_{k}^{(m)}$ is represented by $D_{(\mathcal{P}_{k}^{(m)})}=\sum_{i \in \mathcal{P}_{k}^{(m)}}D_{i}$. For the $k$th diffusion round, the index of the next trainer PUE for model $m$ in $(k-1)$th diffusion round is represented by $i_{k}^{(m)}$, and the following relationship holds:
\begin{equation}
\label{eq:trainer_PUE}
    \mathcal{P}_{k}^{(m)}=\mathcal{P}_{k-1}^{(m)} \cup \{i_{k}^{(m)}\}.
\end{equation}
Here, the entire trainer PUEs in the $k$th diffusion round are represented by the vector $\mathbf{i}_{k}=\{i_{k}^{(1)}, i_{k}^{(2)}, ..., i_{k}^{(M)}\}$.\par

Each PUE computes losses (empirical risks) for its private data batches during local training. Let $l(\mathbf{w};x)$ denote the loss of sample $x$ for training parameter $\mathbf{w}$. The expected loss function of the local model $m$ in the $t$th communication round is the average of the expected losses of PUEs in the diffusion chain $\mathcal{P}_{K_{t}}^{(m)}$, which can be expressed as
\begin{equation}
\label{eq:FedDif_loss}
    \mathcal{J}(\mathbf{w}_{t, K_{t}}^{(m)}) := \frac{1}{\lvert \mathcal{P}_{K_{t}}^{(m)} \rvert}\sum_{i \in \mathcal{P}_{K_{t}}^{(m)}}\mathbb{E}_{x,y \sim P(X_{i})}[l(\mathbf{w}_{t, K_{t}}^{(m)};x)].
\end{equation}
PUEs update the local model with their private data using a stochastic gradient descent (SGD)
\footnote{Each trainer PUE incrementally updates the model by accumulating gradients for its data onto the model's parameters during training. In other words, it is not necessary to calculate gradients for all data at once during the model training within the diffusion chain.}
as
\begin{align}
\label{eq:FedDif_SGD}
    \begin{aligned}
        \mathbf{w}_{t, k}^{(m)} & = \mathbf{w}_{t, k-1}^{(m)} - \eta \triangledown \mathcal{J}(\mathbf{w}_{t, k-1}^{(m)}) \\
        & = \mathbf{w}_{t, k-1}^{(m)} - \frac{\eta}{\lvert\mathcal{P}_{k-1}^{(m)}\rvert} \sum_{i \in \mathcal{P}_{k-1}^{(m)}} \triangledown \mathbb{E}_{x,y \sim P(X_{i})}[l(\mathbf{w}_{t, k-1}^{(m)};x)],
    \end{aligned}
\end{align}
where $\eta$ denotes learning rate. All local models applied the same hyperparameters, such as the learning rate and momentum of SGD sent by the BS.\par
In the case of centralized learning for comparison, the ML model is trained using a universal dataset with IID data that follows the distribution $P(X_{g})$. The loss function with the universal dataset can be expressed by the expectation of the loss function as
\begin{equation}
\label{eq:centralized_loss}
    \mathcal{J}(\mathbf{w}^{(c)}_{t}) := \mathbb{E}_{x,y \sim P(X_{g})}[l(\mathbf{w}^{(c)}_{t};x)],
\end{equation}
where $\mathbf{w}_{t}^{(c)}$ is the parameter of the centralized learning model in $t$th iteration. The BS updates its model based on SGD, which can be expressed as
\begin{align}
\label{eq:centralized_SGD}
    \begin{aligned}
        \mathbf{w}_{t, k}^{(c)} & = \mathbf{w}_{t, k-1}^{(c)} - \eta \triangledown \mathcal{J}(\mathbf{w}_{t, k-1}^{(c)}) \\
        & = \mathbf{w}_{t, k-1}^{(c)} - \eta \triangledown \mathbb{E}_{x,y \sim P(X_{g})}[l(\mathbf{w}_{t, k-1}^{(c)};x)].
    \end{aligned}
\end{align}
Here, $\mathbf{w}_{t, k}^{(c)}$ denotes the parameter after $(k + \sum_{i=1}^{t}K_{i})$ epochs.

After local training and model diffusion, the BS aggregates the entire local model and updates the global model. The BS not only sets the training policy for all local models but also aggregates local models based on FedAvg~\citep{ref:fedavg}, which can be expressed as
\begin{equation}
\label{eq:FedAvg}
    \mathbf{w}_{t}^{(g)} = \sum_{m \in \mathcal{M}}\frac{D_{(\mathcal{P}_{K_{t}}^{(m)})}}{\sum_{m' \in \mathcal{M}}D_{(\mathcal{P}_{K_{t}}^{(m')})}}\mathbf{w}_{t, K_{t}}^{(m)}.
\end{equation}
Based on (\ref{eq:FedAvg}), the BS aggregates the local models by considering the size of the training dataset of PUEs. After sufficient diffusion rounds, the BS obtains a global model whose performance is similar to that of a model trained using the universal IID dataset. This is investigated in Proposition~\ref{prop2} in Section~\ref{section4}.

To improve the communication efficiency of FedDif in wireless networks, we design a communication model to formulate an optimization problem. There are two types of communication: control messages and model transmissions, and we only consider the communication cost of the model transmission. We employ D2D communications
\footnote{In wireless networks, UEs have lower communication capacity than the BS due to their limited transmit power. Consequently, collecting and redistributing all local models by the BS may require more time and frequency resources. FedDif can be compatible with various systems if the communication costs are properly defined.}
with overlying cellular networks in which D2D pairs equally utilize uplink data channels with CUEs for model transmission.
Model transmission may degrade the quality of the data channel in wireless networks because the size of the ML model is large. Therefore, we consider communication costs as the bandwidth required to send model $m$ of $S$ bits in model transmission.\par

Let $g_{i, j}^{(k)}$ denote the channel coefficient from transmitter $i$ to receiver $j$~\citep{ref:channel_modeling, ref:selected1} as
\begin{equation}
    g_{i, j}^{(k)} = \sqrt{\beta_{i, j}^{(k)}}h_{i, j}^{(k)}.
\end{equation}
Here, the wireless channel can be attenuated by channel fading, as the coefficients of small-scale fading $h_{i, j}^{(k)}$ and large-scale fading $\beta_{i, j}^{(k)}$. Small-scale fading is modeled using the Rayleigh fading channel as $h_{i, j}^{(k)} \sim \mathcal{CN}(\boldsymbol{0}, \boldsymbol{I})$. Large-scale fading can be modeled as
\begin{equation}
    \beta_{i, j}^{(k)} = \beta_{0} - 10\kappa\log_{10}\left( \frac{d_{i, j}^{(k)}}{d_{0}} \right),
\end{equation}
where $\beta_{0}$ denotes the large-scale path loss coefficient at a reference distance of $d_{0}$, and $\kappa$ denotes the pathloss exponent~\citep{ref:channel_modeling, ref:selected1}.
We assume an overlay mode of D2D communications, in which D2D pairs and CUEs utilize orthogonal radio resources, for example, orthogonal frequency-division multiple access (OFDMA), to reduce the interference between D2D pairs. We can obtain the spectral efficiency from PUE $i$ to $j$ as
\begin{equation}
    \label{eq:datarate_PUE}
    \gamma_{i, j}^{(k)} = \log\left(1 + \frac{\vert g_{i, j}^{(k)} \vert^{2} p_{i}^{(k)}}{\sigma^{2}}\right),
\end{equation}
where $\sigma^{2} = \sigma_{0}B$ denotes additive white Gaussian noise with power spectral density $\sigma_{0}$. $p_{i}$ denotes the transmit power of PUE $i$.
Finally, we formalize the total bandwidth required to send model $m$ in the $k$th diffusion round as 
\begin{equation}
    \label{eq:required_BW}
    B_{k}^{(m)}=\frac{S}{\gamma_{i_{k-1}^{(m)}, i_{k}^{(m)}}^{(k)}},
\end{equation}
where $S$ is the size of the ML model. Here, PUE $i_{k-1}^{(m)}$ sends model $m$ to PUE $i_{k}^{(m)}$ during $k$-th diffusion round.\par

\subsection{Representations of Data Distribution}\label{section3B}
We define the representation of the data distribution of a PUE as the data state information (DSI) vector $\mathbf{d}_{i} \sim \mathcal{D}_{i}$. The DSI vector is probabilistic in that each component is bounded in the closed interval of $[0, 1]$ and the components sum to one. Moreover, we represent the cumulative data distribution of model $m$ after $k$ diffusion iterations, called the degree of learning (DoL), as
\begin{equation}
    \label{eq:DoL}
    \psi_{k}^{(m)}=\frac{1}{D_{(\mathcal{P}_{k}^{(m)})}}\left(D_{(\mathcal{P}_{k-1}^{(m)})}\psi_{k-1}^{(m)}+D_{i_{k}^{(m)}}\mathbf{d}_{i_{k}^{(m)}}\right),
\end{equation}
where $\mathbf{d}_{i_{k}^{(m)}}$ is the DSI of the PUE $i_{k}^{(m)}$. The DoL $\psi_{k}^{(m)}$ of model $m$ at the $k$th diffusion indicates the cumulative ratio of the learned data size for each class in all PUEs of the diffusion subchain $\mathcal{P}_{k}^{(m)}$ and follows all the properties of the DSI.
Note that there is no privacy leakage because constructing the diffusion chain does not use the DSI of the user but instead uses the DoL of the model. Privacy could only be compromised if the DoL in the first diffusion round is identical to the DSI of the first trainer PUE. This problem is solved by adding Gaussian noise to the DoL during the first round of diffusion.
We assume that the DoL of the model trained in a centralized manner with IID data follows a uniform distribution, that is, $\mathbf{I}=\frac{1}{C}\mathbf{1}$ where $\mathbf{1}$ is a vector of all ones.\par

Let $\delta(\cdot)$ define the \emph{IID distance}, which measures the distance between distributions of a specific probability $P$ and uniform. Wasserstein-1, called the Earth-mover distance, is widely used to measure the distance between two distributions as follows:
\begin{equation}
    \label{eq:wdist}
    W_{1}(P_{a}, P_{b}) = \inf_{\gamma\in\Pi\left(P_{a}, P_{b}\right)}\mathbb{E}_{\left(x, y\right) \sim \gamma}[\left\lVert x-y \right\rVert],
\end{equation}
where $\Pi\left(P_{a}, P_{b}\right)$ denotes the set of joint distributions of two distributions $P_{a}$ and $P_{b}$~\citep{ref:WGAN}. Based on (\ref{eq:wdist}), we quantify IID distance of the model $m$ in the $k$th diffusion as follows:
\begin{align}
\label{eq:IID_dist}
    \begin{aligned}
        & W_{1}\left(\psi_{k}^{(m)}, \mathcal{U}\right) \\
        & = \bigg\Vert \frac{1}{D_{(\mathcal{P}_{k}^{(m)})}}\left(D_{(\mathcal{P}_{k-1}^{(m)})}\psi_{k-1}^{(m)}+D_{i_{k}^{(m)}}\mathbf{d}_{i_{k}^{(m)}}\right) - \frac{1}{C}\mathbf{1} \bigg\Vert,
    \end{aligned}
\end{align}
where $i_{k}^{(m)}$ and $\psi_{k}^{(m)}$ are the trainer PUE and DoL of model $m$ in the $k$th diffusion, respectively. $D_{i_{k}^{(m)}}$ and $\mathbf{d}_{i_{k}^{(m)}}$ respectively denote the data size and DSI of the PUE $i_{k}^{(m)}$. $\frac{1}{C}\mathbf{1}_{C}$ denotes the probability vector of the uniform distribution $\mathcal{U}$ where $\mathbf{1}_{C}=[1, 1, ..., 1] \in \mathbb{R}^{C}$ is $C$-dimensional one-vector.\par

\subsection{Problem formulation}\label{section3C}
There is a trade-off between the communication cost of diffusion and the learning performance of the global model. For example, FedDif requires more communication resources than typical FL in the short term. However, in the long term, the entire number of iterations required to obtain the required performance of the global model may decrease because the optimization trajectory of the global model after diffusion can become much closer to the optimal trajectory than the typical FL. FedDif aims to coordinate the trade-off by determining the optimal diffusion chain whose PUEs maximize the variance of IID distance and minimize the communication cost.\par
Two primary variables are a set of the next trainer PUEs $i_{k}^{(m)}$ and the required communication resources as $\mathbf{i}_{k} = \{i_{k}^{(m)}: m \in \mathcal{M}\} \in \mathbb{R}^{M}$ and $\mathbf{B}_{k} = \{B_{k}^{(m)}: m \in \mathcal{M}\} \in \mathbb{R}^{M}$, respectively. Note that $B_{k}^{(m)}$ denotes the total bandwidth required for model $m$ in the $k$th diffusion round. We then can define the \emph{diffusion efficiency}, which indicates the variance of IID distance against the total bandwidth required to diffuse the entire model in the $k$th diffusion round, as
\begin{equation}
\label{eq:diffusion_efficiency}
    E(\textbf{i}_{k}, \textbf{B}_{k}) = \frac{1}{M}\sum_{m \in \mathcal{M}}\frac{\delta_{i_{k}^{(m)}}}{B_{k}^{(m)}}.
\end{equation}
Here, $\delta_{i_{k}^{(m)}}$ indicates the variance in IID distance when the next trainer PUE $i_{k}^{(m)}$ in the $k$th diffusion round trains model $m$ as follows:
\begin{equation}
\label{eq:improvement_dif_eff}
    \delta_{i_{k}^{(m)}} = W_{1}\left(\psi_{k-1}^{(m)}, \mathcal{U}\right) - W_{1}\left(\psi_{k}^{(m)}, \mathcal{U}\right),
\end{equation}
where $\psi_{k}^{(m)}$ is determined by PUE $i_{k}^{(m)}$. Note that the variance in IID distance decides the sign of the diffusion efficiency. For example, the variance in IID distance is positive if the IID distance decreases, which means that the DoL approaches a uniform distribution. Moreover, diffusion efficiency may decrease when communication resources are insufficient for diffusing models or the variance is trivial against the required resources.\par
Therefore, we can conclude that FedDif is an algorithm for solving the diffusion efficiency maximization problem by determining the optimal $\textbf{i}_{k}^{*}$ and $\textbf{B}_{k}^{*}$ as follows:
\begin{subequations}
    \label{eq:objFunc}
    \begin{alignat}{3}
        \max_{\textbf{i}_{k}, \textbf{B}_{k}} \; &E(\textbf{i}_{k}, \textbf{B}_{k}) \label{eq:objFuncA} \\
        \textrm{s.t.} \quad &\delta_{i_{k}^{(m)}} \geq 0, \hfill \forall i_{k}^{(m)} \in \textbf{i}_{k}, \label{eq:ConstB} \\
        &i_{k}^{(m)} \notin \mathcal{P}_{k-1}^{(m)}, \hfill \forall i_{k}^{(m)} \in \textbf{i}_{k}, \label{eq:ConstC} \\
        &i_{k}^{(m_{1})} \neq i_{k}^{(m_{2})}, \forall i_{k}^{(m_{1})}, i_{k}^{(m_{2})} \in \textbf{i}_{k}, \label{eq:ConstD} \\
        &\gamma_{i_{k-1}^{(m)}, i_{k}^{(m)}}^{(k)} \geq \gamma_{\textrm{min}}, \hfill \forall i_{k}^{(m)} \in \textbf{i}_{k}, \label{eq:ConstE} \\
        &\sum_{m \in \mathcal{M}}B_{k}^{(m)} \leq B_{k} - \sum_{i \in \mathcal{N}_{C}}\Tilde{B}_{k}^{(i)}, \hfill \forall B_{k}^{(m)} \in \textbf{B}_{k}. \label{eq:ConstF}
    \end{alignat}
\end{subequations}
Here, constraint~\eqref{eq:ConstB} ensures that FedDif does not deteriorate the diffusion efficiency. Retraining the model is prohibited by the constraint~\eqref{eq:ConstC} because it may induce weight divergence. Overtraining a model using a single dataset can cause a weight-divergence problem~\citep{ref:noniid1}. The constraint~\eqref{eq:ConstD} indicates that the PUEs can train only one model in a diffusion round. Constraint~\eqref{eq:ConstE} indicates the minimum tolerable QoS requirement of PUEs. D2D communications overlying cellular networks are employed for the diffusion mechanism, represented by the constraint~\eqref{eq:ConstF}.\par
It can be easily seen that the diffusion efficiency maximization problem in~\eqref{eq:objFunc} is a combinatorial optimization problem. It is difficult to obtain a solution directly because the set of feasible solutions is discrete. Therefore, based on auction theory, we design a diffusion strategy to find a feasible solution that simultaneously minimizes the IID distance and required spectral resources. In the following section, we provide a theoretical basis for feasible solutions that can solve the weight-divergence problem and minimize IID distance.\par

\section{Communication-efficient diffusion strategy}\label{section5}
In this section, we introduce a novel training strategy called FedDif, which selects neighboring PUEs to maximize diffusion efficiency and accumulates the data distributions of different PUEs by the diffusion mechanism. FedDif aims to accumulate different non-IID data distributions in local models before global aggregation.\par

FedDif includes four diffusion steps before global aggregation: \emph{DoL broadcasting}, \emph{candidates for IID distance reporting}, \emph{diffusion configuration}, and \emph{model transmission}, as illustrated in Fig.~\ref{fig:diffusion_concept}. The entire procedure of FedDif is described in Algorithm~\ref{alg:dif_mechanism}. PUEs advertise the DoL $\psi_{k-1}^{(m)}$ of the responsible model $m$ to their neighboring PUEs (\emph{\textbf{DoL broadcasting}})~\footnote{The data privacy of PUEs is preserved because the DoL reflects the cumulative distribution of learned data, and adversaries cannot determine the DSI of PUEs solely from the DoL. Although the DoL in the first diffusion round may indicate the DSI of the first trainer, we prevent privacy leakage by adding noise to the first DoL. Furthermore, the noisy DoL in the first diffusion round does not affect the convergence of the IID distance, as demonstrated in Proposition~\ref{prop2} in Section~\ref{section4}.}. 
Then, neighboring PUEs calculate the candidate DoL of the $k$th diffusion round for all received DoLs of the previous diffusion round using their DSI and report the candidates of IID distance to the BS (\emph{\textbf{candidates of IID distance reporting}}). Candidates for the IID distance indicate the estimated IID distance for the next diffusion round. The BS sends scheduling policies to PUEs for model transmission (\emph{\textbf{diffusion configuration}}). Finally, PUEs send their model to the next trainer PUEs via the scheduled data channel (\emph{\textbf{model transmission}}). We provide the basis for how FedDif can address the performance degradation problem of FL owing to non-IID data in Section~\ref{section5}.\par

Although FedDif enhances the learning performance of FL, it may become a critical issue if it requires additional communication resources for diffusion. The diffusion mechanism could consume excessive communication resources when PUEs send local models to minimize the IID distance without considering channel conditions.
\begin{algorithm}[H]
    \caption{FedDif}
    \label{alg:dif_mechanism}
    \textbf{Input:} $\eta$: hyperparameter, $\mathcal{N}_{P}$: a set of PUEs, $\mathcal{M}$: a set of local models, $T$: maximum communication rounds, $\varepsilon$: minimum tolerable IID distance. \\ 
    \textbf{Output:} $\mathbf{w}_{T}^{(g)}$: parameters of the global model at $T$ communication round.
    \begin{algorithmic}[1]
        \State Initialize the global model $\mathbf{w}_{0}^{(g)}$ and broadcast the hyper-parameters
        \For{$t=1, 2, ..., T$}
            \State BS broadcasts $\mathcal{M}$ to PUEs.
            \State $k \gets 0$, 
            \State $\mathcal{P}_{k}^{(m)} \gets \emptyset$, 
            \State $D_{(\mathcal{P}_{k}^{(m)})} \gets 0$, 
            \State $\psi_{k}^{(m)} \gets \mathbf{0}$
            \For{$m \in \mathcal{M}$}
                \State \multiline{Trainer PUE $i_{k}^{(m)}$ of the model $m$ forms the local dataset $\mathcal{D}_{i_{k}^{(m)}}$ with the DSI $\mathbf{d}_{i_{k}^{(m)}}$.}
                \State $\mathbf{w}_{t, k}^{m}, \mathcal{P}_{k}^{m} \gets$ \text{LocalUpdate}{($k$, $m$, $i_{k}^{m}$)}
            \EndFor
            \While{$W_{1}(\psi_{k}^{(m)}, \mathcal{U}) > \varepsilon$}
                \State $k \gets k+1$
                \For{$i \in \mathcal{N}_{p}$}
                    \State Broadcast $\psi_{k-1}^{(m)}$ to the other PUEs.
                    \State Calculate valuation $v_{i,k}^{(m)} \in \textbf{bid}_{k}^{(m)}$.
                    \State Transmit $\textbf{bid}_{k}^{(m)}$ and $\textbf{g}_{k}^{(m)}$ to BS.
                \EndFor
                \State \multiline{BS determines $\mathbf{i}_{k}^{*}$ and $\mathbf{B}_{k}^{*}$ by winner selection algorithm and schedules PUEs.}
                \State Scheduled PUEs transmit their local model.
                \For{$m \in \mathcal{M}$}
                    \State $\mathbf{w}_{t, k}^{m}, \mathcal{P}_{k}^{m} \gets$ \text{LocalUpdate}{($k$, $m$, $i_{k}^{m}$)}
                \EndFor
            \EndWhile
            \State $\mathbf{w}^{g}_{t} \gets \sum_{m \in \mathcal{M}}\frac{D_{(\mathcal{P}_{k}^{m})}}{\sum_{m' \in \mathcal{M}}D_{(\mathcal{P}_{k}^{m'})}}\mathbf{w}^{m}_{t, k}$
        \EndFor
        \Function {LocalUpdate} {$k$, $m$, $i_{k}^{m}$}
            \State $\mathcal{P}_{k}^{m} \gets \mathcal{P}_{k-1}^{m} \cup \{i_{k}^{m}\}$
            \State $D_{(\mathcal{P}_{k}^{m})} \gets D_{(\mathcal{P}_{k-1}^{m})} + D_{i_{k}^{m}}$ 
            \State $\psi_{k}^{m} \gets \frac{1}{D_{(\mathcal{P}_{k}^{m})}}(D_{(\mathcal{P}_{k-1}^{m})}\psi_{k-1}^{m} + D_{i}\mathbf{d}_{i_{k}^{m}})$
            \State $\mathbf{w}_{t, k}^{m} \gets \mathbf{w}_{t, k-1}^{m} - \eta\triangledown\mathbb{E}_{x, y \sim P(X_{i_{k}^{m}})}[l(\mathbf{w}_{t, k-1}^{m};x)]$
            \Return $\mathbf{w}_{t, k}^{m}$, $\mathcal{P}_{k}^{m}$
        \EndFunction
    \end{algorithmic}
\end{algorithm}
For example, PUEs may transmit the model to maximize the learning performance despite poor channel conditions. This can degrade the communication quality for all UEs in the network and prolong the overall training time. Therefore, it is essential to design a communication-efficient diffusion strategy, and auction theory is suitable for determining the optimal point between learning performance and communication costs and competitively achieving all user objectives without sharing private data~\citep{ref:auction1, ref:auction2}.\par



\subsection{Modeling the bidding price}\label{section5A}
We assume a second-price auction in which the bidder with the highest bidding price becomes the winner and pays the second-highest price. PUEs should rationally minimize the IID distance of their local model because paying the original valuation of each PUE is a dominant strategy. PUEs set the valuation as the expected variance of IID distance when training the models of neighboring PUEs with their local data. Based on \eqref{eq:improvement_dif_eff}, we can formulate the valuation of model $m$ for PUE $i$ using the variance of IID distance as follows:
\begin{equation}
\label{eq:valuation}
    v_{i, k}^{(m)} = W_{1}(\psi_{k-1}^{(m)}, \mathcal{U}) - W_{1}(\Tilde{\psi}_{i, k}^{(m)}, \mathcal{U}),
\end{equation}
where we define the candidate of DoL for the PUE $i$ as $\Tilde{\psi}_{i, k}^{(m)} \triangleq \frac{1}{D_{(\mathcal{P}_{k}^{(m)})}}(D_{(\mathcal{P}_{k-1}^{(m)})}\psi_{k-1}^{(m)}+D_{i}\mathbf{d}_{i})$. The bidding price for transmitting model $m$ comprises the valuations of every PUE and can be expressed as
\begin{equation}
\label{eq:bid}
    \begin{aligned}
        \textbf{bid}_{k}^{(m)} = \left[v^{(m)}_{1,k}, ..., v^{(m)}_{N_{P},k} \right].
    \end{aligned}
\end{equation}
In addition, the BS should collect channel state information between PUEs to obtain the required bandwidth for model transmission. With the bidding prices, PUEs also send a bundle of channel state information $\textbf{g}_{k}^{(m)}$ between the previous trainer PUE $i_{k-1}^{(m)}$ and the other PUEs to the BS. A bundle of channel state information can be expressed as
\begin{equation}
    \label{eq:csi}
    \begin{aligned}
        \textbf{g}_{k}^{(m)} = \left[ g_{i_{k-1}^{(m)},1}, ..., g_{i_{k-1}^{(m)},N_{P}} \right].
    \end{aligned}
\end{equation}

\subsection{Winner selection algorithm}\label{section5B}
After sending the bids to the BS, it determines the auction winner. The BS can achieve the maximum variance in IID distance when it selects PUEs with the highest bidding price as the auction winner. However, PUEs require more bandwidth to send to the model owing to poor channel conditions. Moreover, errors may occur in the model because model transmissions through poor channels may increase the bit error rate. Thus, the BS should select the winner PUEs by considering the diffusion efficiency, which includes the IID distance and required bandwidth for the model transmission.

\begin{algorithm}[t]
    \caption{Winner selection algorithm}
    \label{alg:winner_selection}
    \textbf{Input:} $\mathcal{M}$, $\mathcal{N}_{P}$, $\left[ \textbf{bid}^{(1)}_{k}, ..., \textbf{bid}^{(M)}_{k}\right]$, $\left[\textbf{g}^{(1)}_{k}, ..., \textbf{g}^{(M)}_{k}\right]$\\   
    \textbf{Output:} $\mathbf{i}_{k}^{*}$, $\mathbf{B}^{*}_{k}$  
    \begin{algorithmic}[1]
        \State Construct a set of edges $(m, i) \in \mathcal{E}$, $\forall m \in \mathcal{M}, i \in \mathcal{N}_{P}$
        \State Calculate each edge weight $c(m, i)$ by \eqref{eq:weight_hungarian}
        \State Construct a bipartite graph $\mathcal{G} = (\mathcal{M}, \mathcal{N}_{P}, \mathcal{E})$
        \State Find the maximal matching $\mathcal{R}^{*} \subseteq \mathcal{M} \times \mathcal{N}_{P}$ by Kuhn- Munkres algorithm
        \State Select auction winners $\mathbf{i}_{k}^{*} = \left[\left\{ i : (m, i) \in \mathcal{R}^{*} \right\}\right]$
        \State Allocate communication resources \par $\mathbf{B}^{*}_{k} = \left[\left\{ \Tilde{B}^{(m)}_{i, k} : (m, i) \in \mathcal{R}^{*} \right\}\right]$
    \end{algorithmic}
\end{algorithm}

We employ the Kuhn–Munkres algorithm, which can find the optimal matching in polynomial time, to match the models and next trainer PUEs by considering the diffusion efficiency. Let $\mathcal{G}$ denote a bipartite graph representing the relationship between a set of workers and job vertices by using a set of edges. By applying our problem, we can represent $\mathcal{G} = (\mathcal{M}, \mathcal{N}_{P}, \mathcal{E})$, where a set of worker and job vertices can be denoted by $\mathcal{M}$ and $\mathcal{N}_{P}$, respectively. $\mathcal{E}$ denotes a set of edges, where each edge can be represented by a pair of the model and next trainer PUE as $(m, i) \in \mathcal{E}$. Then, the Kuhn-Munkres algorithm aims to find the optimal set of edges for which the sum of the weights is maximized. The weight comprises the valuation and required communication resources, expressed as
\begin{equation}
    \label{eq:weight_hungarian}
    c(m, i) = 
    \begin{cases}
        v_{i, k}^{(m)} / \Tilde{B}_{i, k}^{(m)}, & \mbox{if }\eqref{eq:ConstB}, \eqref{eq:ConstC}, \mbox{and } \eqref{eq:ConstE} \mbox{ hold}\\
        0, & \mbox{otherwise},
    \end{cases}
\end{equation}
where $v_{i, k}^{(m)} \in \textbf{bid}_{k}^{(m)}$ and $\Tilde{B}_{i, k}^{(m)}$ denote the valuation and communication resources required for PUE $i$ to transmit model $m$, respectively. The required communication resource can be expressed as
\begin{equation}
    \label{eq:comm_resource}
    \Tilde{B}_{i, k}^{(m)} = S / \Tilde{\gamma}_{i_{k-1}^{(m)}, i}^{(k)},
\end{equation}
where $\Tilde{\gamma}_{i_{k-1}^{(m)}, i}^{(k)}$ denotes the expected spectral efficiency when the previous trainer PUE $i_{k-1}^{(m)}$ sends model $m$ to PUE $i$. Based on \eqref{eq:weight_hungarian}, we can obtain the feasible solutions to the diffusion efficiency maximization problem \eqref{eq:objFunc} by finding the maximal matching using the Kuhn-Munkres algorithm. The winner selection algorithm is described in Algorithm~\ref{alg:winner_selection}.

\subsection{Complexity analysis}\label{section5D}
FedDif operates in $K$ diffusion rounds, and the winner selection algorithm is performed at each. In the worst case, the maximum number of diffusion rounds is $\mathcal{O}\left(N_{P}(N_{P}-1)\right)$, where only one PUE transmits the model at each diffusion round until every model learns the data distribution of all PUEs. In each diffusion round, bipartite graph construction, Kuhn–Munkres algorithm, and resource allocation based on the first-come-first-served (FCFS) algorithm are performed. To construct a bipartite graph, the BS calculates the weight of each edge with complexity $\mathcal{O}(MN_{P})$. The Kuhn-Munkres algorithm follows the complexity $\mathcal{O}(\max(M, N_{P})^{3})$ as widely known. $\max(M, N_{P}) = N_{P}$ holds because the BS can initiate local models for most PUEs in the global initialization phase. The complexity of FCFS based on the greedy algorithm is $\mathcal{O}(N_{P})$. Consequently, the computational complexity of FedDif can be represented as 
\begin{equation}
    \label{eq:entire_compleity}
    \mathcal{O}\left({N_{P}(N_{P} - 1)}\left(MN_{P} + {N_{P}}^{3}+N_{P}\right) \right) = \mathcal{O}\left({N_{P}}^{5}\right).
\end{equation}
However, the complexity of FedDif could be lower than \eqref{eq:entire_compleity} by the minimum tolerable QoS $\gamma_{\textrm{min}}$ and diffusion efficiency $\varepsilon$ in practice. Moreover, the complexity can vary depending on the complexity of the resource allocation algorithm.

\section{Theoretical analysis}\label{section4}
In this section, we provide a theoretical analysis of FedDif, which can reduce biases in non-IID data and improve learning performance. We first investigate the relationship between the learning performance and the probability distance of the user data. Then, we prove that FedDif can reduce the probability distance when PUEs have sufficiently diffuse the models. Based on the proofs, we provide remarks on how FedDif can mitigate weight divergence with sufficient diffusion.\par

We consider two assumptions that have been widely adopted in existing FL~\citep{ref:assumption1, ref:cefl5, ref:noniid1, ref:noniid3, ref:selected1}.
\begin{assumption}[Lipschitz continuity]
\label{assumption1}
    \emph{For any $\mathbf{w}_{1}$ and $\mathbf{w}_{2}$, if the loss function $l(\mathbf{w}; x)$ is derivable and convex, the gradient of $l(\mathbf{w}; x)$ is $\lambda$-Lipschitz continuous as:}
    \begin{equation}
    \label{eq:lipschitz}
        \left\langle \triangledown l(\mathbf{w}_{1}; x) - \triangledown l(\mathbf{w}_{2}; x), \mathbf{w}_{1} - \mathbf{w}_{2} \right\rangle \leq \lambda {\lVert \mathbf{w}_{1} - \mathbf{w}_{2} \rVert}^{2},
    \end{equation}
    \emph{where $\left\langle \cdot \right\rangle$ represents the inner product and $\lambda$ is a Lipschitz constant with $\lambda > 0$.}
\end{assumption}
\begin{assumption}[Upper bound of gradients]
\label{assumption2}
    \emph{A fundamental cause of weight divergence is the gradient exploding. For any $\mathbf{w}$, the upper bound of $\triangledown l(\mathbf{w}; x)$ to restrict gradient explosion can be expressed as:}
    \begin{equation}
    \label{eq:upper_bound_gradient}
        \left\lVert \triangledown l(\mathbf{w}; x) \right\rVert \leq \mu,
    \end{equation}
    \emph{where $\mu$ is an upper bound of the gradient.}
\end{assumption}
The assumptions may be satisfied only for ML algorithms whose loss function is convex, such as support vector machine (SVM) and logistic regression. Although the loss function of a deep neural network (DNN) is non-convex, we empirically demonstrate that FedDif is still effective for DNN in Section \ref{section6}.\par

The optimal points of the local models are different because each user has a unique optimization trajectory for their private dataset. The aggregation of the local models optimized in different directions prevents the global model from determining the global optimum. On the other hand, training the local models in a similar optimization direction can mitigate the performance degradation of the global model. We attempt to induce a similar trajectory of local models by spreading them to learn different non-IID data distributions before global aggregation. Thus, we prove that the diffusion mechanism can approximate the global model to a model trained with IID data.\par

For simplicity, we use $K$ as the last diffusion round of $t$th communication round $K_{t}$. We first define the expected weight difference $\Delta_{t}$ in the $t$th communication round based on FedAvg as follows:
\begin{equation}
\label{eq:weight_divergence}
    \begin{aligned}
        \Delta_{t} & = \mathbb{E}\left[\left\lVert \mathbf{w}_{t}^{(g)} - \mathbf{w}_{t, K}^{(c)} \right\rVert\right] \\
        & = \mathbb{E}\left[\left\lVert \sum_{m \in \mathcal{M}}\frac{D_{\left(\mathcal{P}_{K}^{(m)}\right)}}{\sum_{m' \in \mathcal{M}}D_{\left(\mathcal{P}_{K}^{(m')}\right)}}\mathbf{w}_{t, K}^{(m)} - \mathbf{w}_{t, K}^{(c)} \right\rVert \right] \\
        & \leq \mathbb{E}\left[\sum_{m \in \mathcal{M}}\frac{D_{\left(\mathcal{P}_{K}^{(m)}\right)}}{\sum_{m' \in \mathcal{M}}D_{\left(\mathcal{P}_{K}^{(m')}\right)}}\left\lVert \mathbf{w}_{t, K}^{(m)} - \mathbf{w}_{t, K}^{(c)} \right\rVert \right].
    \end{aligned}
\end{equation}
We can clearly observe that the weight difference between the local and centralized learning models can be attributed to weight divergence. To reveal the cause and extent of weight divergence, the upper bound of the weight difference between the local model $m$ and the centralized learning model is provided by Proposition~\ref{prop1}.

\begin{prop}
\label{prop1}
    \emph{An upper bound of weight difference can be expressed as}
    \begin{equation}
        \label{eq:prop1}
        \begin{aligned}
            & \lVert \mathbf{w}_{t, K}^{(m)} - \mathbf{w}_{t, K}^{(c)} \rVert \leq (a^{(m)})^{K} \lVert \mathbf{w}_{t, 0}^{(m)} - \mathbf{w}^{(c)}_{t, 0} \rVert \\
            & + \frac{(a^{(m)})^{K} - 1}{a^{(m)} - 1} \frac{\eta\mu}{\lvert \mathcal{P}_{K}^{(m)}\rvert} \sum_{i \in \mathcal{P}_{K}^{(m)}}\sum_{c=1}^{C} \lVert P(X_{i}=c) - P(X_{g}=c) \rVert,
        \end{aligned}
    \end{equation}
    \emph{where $a^{(m)} = 1 + \frac{\eta}{\lvert\mathcal{P}_{K}^{(m)}\rvert} \sum_{i \in \mathcal{P}_{K}^{(m)}}\lambda_{i}$.} \\
    \begin{proof}
        Based on (\ref{eq:centralized_SGD}) and (\ref{eq:FedDif_SGD}), we can rewrite (\ref{eq:weight_divergence}) as
    \begin{equation}
    \label{pf1:01}
        \begin{aligned}
            & \lVert \mathbf{w}_{t, K}^{(m)} - \mathbf{w}_{t, K}^{(c)} \rVert \\
            & = \bigg\Vert \mathbf{w}_{t, K-1}^{(m)} - \frac{\eta}{\lvert \mathcal{P}_{K-1}^{(m)}\rvert} \sum_{i \in \mathcal{P}_{K-1}^{(m)}} \triangledown \mathbb{E}_{x,y \sim P(X_{i})}[l(\mathbf{w}_{t, K-1}^{(m)};x)] \\
            & \qquad - \mathbf{w}_{t, K-1}^{(c)} + \eta \triangledown \mathbb{E}_{x,y \sim P(X_{g})}[l(\mathbf{w}_{t, K-1}^{(c)};x)] \bigg\Vert \\
            & \leq \lVert \mathbf{w}_{t, K-1}^{(m)} - \mathbf{w}_{t, K-1}^{(c)} \rVert \\
            & \qquad + \frac{\eta}{\lvert \mathcal{P}_{K-1}^{(m)}\rvert} \bigg\Vert \sum_{i \in \mathcal{P}_{K-1}^{(m)}} \Big( \triangledown \mathbb{E}_{x,y \sim P(X_{i})}[l(\mathbf{w}_{t, K-1}^{(m)};x)] \\
            & \qquad \qquad \qquad - \triangledown \mathbb{E}_{x,y \sim P(X_{g})}[l(\mathbf{w}_{t, K-1}^{(c)};x)] \Big) \bigg\Vert.
        \end{aligned}
    \end{equation}
    A set of PUEs can be divided into diffusion subchain and their complements, that is, $\mathcal{N}_{p}=\mathcal{P}_{k-1}^{(m)} \cup \bar{\mathcal{P}}_{k-1}^{(m)}$. The random variable of the universal dataset can be expressed as $X_{g}=\cup_{i \in \mathcal{N}_{p}}X_{i}$.
    According to the triangular inequality, we can rewrite \eqref{pf1:01} as
    \begin{align*}
        \begin{aligned}
            & \lVert \mathbf{w}_{t, K-1}^{(m)} - \mathbf{w}_{t, K-1}^{(c)} \rVert \\
            & \quad + \frac{\eta}{\lvert \mathcal{P}_{K-1}^{(m)}\rvert} \bigg\Vert \sum_{i \in \mathcal{P}_{K-1}^{(m)}} \Big( \triangledown \mathbb{E}_{x,y \sim P(X_{i})}[l(\mathbf{w}_{t, K-1}^{(m)};x)] \\
            & \qquad - \triangledown \mathbb{E}_{x,y \sim P(X_{g})}[l(\mathbf{w}_{t, K-1}^{(c)};x)] \Big) \bigg\Vert \\
            & = \lVert \mathbf{w}_{t, K-1}^{(m)} - \mathbf{w}_{t, K-1}^{(c)} \rVert \\
            & \quad + \frac{\eta}{\left\vert \mathcal{P}_{K-1}^{(m)}\right\vert} \bigg\Vert \sum_{i \in \mathcal{P}_{K-1}^{(m)}} \Big( \triangledown \mathbb{E}_{x,y \sim P(X_{i})}[l(\mathbf{w}_{t, K-1}^{(m)};x)] \\
            & \qquad - \triangledown \mathbb{E}_{x,y \sim P(X_{i})}[l(\mathbf{w}_{t, K-1}^{(c)};x)] \Big) \\
            & \quad + \Big( \triangledown \mathbb{E}_{x,y \sim P(X_{i})}[l(\mathbf{w}_{t, K-1}^{(c)};x)] \\
            & \qquad - \triangledown \mathbb{E}_{x,y \sim P(X_{g})}[l(\mathbf{w}_{t, K-1}^{(c)};x)] \Big) \bigg\Vert \\
            & \leq \lVert \mathbf{w}_{t, K-1}^{(m)} - \mathbf{w}_{t, K-1}^{(c)} \rVert \\
            & \quad + \frac{\eta}{\lvert \mathcal{P}_{K-1}^{(m)}\rvert} \bigg\{ \sum_{i \in \mathcal{P}_{K-1}^{(m)}} \bigg\Vert \triangledown \mathbb{E}_{x,y \sim P(X_{i})}[l(\mathbf{w}_{t, K-1}^{(m)};x)] \\
        \end{aligned}
    \end{align*}
    \begin{equation}
        \label{pf1:02}
        \begin{aligned}
            & \qquad - \triangledown \mathbb{E}_{x,y \sim P(X_{i})}[l(\mathbf{w}_{t, K-1}^{(c)};x)] \bigg\Vert \\
            & \quad + \sum_{i \in \mathcal{P}_{K-1}^{(m)}} \bigg\Vert \triangledown \mathbb{E}_{x,y \sim P(X_{i})}[l(\mathbf{w}_{t, K-1}^{(c)};x)] \\
            & \qquad - \triangledown \mathbb{E}_{x,y \sim P(X_{g})}[l(\mathbf{w}_{t, K-1}^{(c)};x)] \bigg\Vert \bigg\}.
        \end{aligned}
    \end{equation}
    Based on Assumptions \ref{assumption1}, \ref{assumption2}, and the triangular inequality, we can derive \eqref{pf1:02} as
    \begin{equation}
    \label{pf1:03}
        \begin{aligned}
            & \lVert \mathbf{w}_{t, K-1}^{(m)} - \mathbf{w}_{t, K-1}^{(c)} \rVert \\
            & \quad + \frac{\eta}{\lvert \mathcal{P}_{K-1}^{(m)} \rvert} \bigg\{ \sum_{i \in \mathcal{P}_{K-1}^{(m)}} \bigg\Vert \triangledown \mathbb{E}_{x,y \sim P(X_{i})}[l(\mathbf{w}_{t, K-1}^{(m)};x)] \\
            & \qquad - \triangledown \mathbb{E}_{x,y \sim P(X_{i})}[l(\mathbf{w}_{t, K-1}^{(c)};x)] \bigg\Vert \\
            & \quad + \sum_{i \in \mathcal{P}_{K-1}^{(m)}} \bigg\Vert \triangledown \mathbb{E}_{x,y \sim P(X_{i})}[l(\mathbf{w}_{t, K-1}^{(c)};x)] \\
            & \qquad - \triangledown \mathbb{E}_{x,y \sim P(X_{g})}[l(\mathbf{w}_{t, K-1}^{(c)};x)] \bigg\Vert \bigg\} \\
            & \leq \lVert \mathbf{w}_{t, K-1}^{(m)} - \mathbf{w}_{t, K-1}^{(c)} \rVert \\
            & \quad + \frac{\eta}{\lvert \mathcal{P}_{K-1}^{(m)}\rvert} \Bigg\{ \sum_{i \in \mathcal{P}_{K-1}^{(m)}} \lambda_{i} \lVert \mathbf{w}_{t, K-1}^{(m)} - \mathbf{w}_{t, K-1}^{(c)} \rVert \\
            & \qquad + \sum_{i \in \mathcal{P}_{K-1}^{(m)}} \bigg\Vert \triangledown \mathbb{E}_{x,y \sim P(X_{i})}[l(\mathbf{w}_{t, K-1}^{(c)};x)] \\
            & \qquad - \triangledown \mathbb{E}_{x,y \sim P(X_{g})}[l(\mathbf{w}_{t, K-1}^{(c)};x)] \bigg\Vert \bigg\} \\
            & \leq \lVert \mathbf{w}_{t, K-1}^{(m)} - \mathbf{w}_{t, K-1}^{(c)} \rVert + \frac{\eta}{\lvert \mathcal{P}_{K-1}^{(m)}\rvert} \Bigg\{{\sum_{i \in \mathcal{P}_{K-1}^{(m)}}} \lambda_{i} \lVert \mathbf{w}_{t, K-1}^{(m)} - \mathbf{w}_{t, K-1}^{(c)} \rVert \\
            & + {\sum_{i \in \mathcal{P}_{K-1}^{(m)}}} \sum_{x \in \mathcal{D}_{i}} \sum_{c=1}^{C}  \bigg\Vert \triangledown l(\mathbf{w}_{t, K-1}^{(c)};x) \left(P(X_{i}=c) - P(X_{g}=c)\right) \bigg\Vert \bigg\} \\
            & \leq \lVert \mathbf{w}_{t, K-1}^{(m)} - \mathbf{w}_{t, K-1}^{(c)} \rVert + \frac{\eta}{\lvert \mathcal{P}_{K-1}^{(m)}\rvert} \bigg\{{\sum_{i \in \mathcal{P}_{K-1}^{(m)}}} \lambda_{i} \lVert \mathbf{w}_{t, K-1}^{(m)} - \mathbf{w}_{t, K-1}^{(c)} \rVert \\
            & + \mu \sum_{i \in \mathcal{P}_{K-1}^{(m)}} \sum_{x \in \mathcal{D}_{i}} \sum_{c=1}^{C} \bigg\Vert P(X_{i}=c) - P(X_{g}=c) \bigg\Vert \bigg\} \\
            & = \bigg( 1 + \frac{\eta}{\lvert \mathcal{P}_{K-1}^{(m)}\rvert} \sum_{i \in \mathcal{P}_{K-1}^{(m)}} \lambda_{i} \bigg) \lVert \mathbf{w}_{t, K-1}^{(m)} - \mathbf{w}_{t, K-1}^{(c)} \rVert \\
            & + \frac{\eta\mu}{\lvert \mathcal{P}_{K-1}^{(m)}\rvert} \sum_{i \in \mathcal{P}_{K-1}^{(m)}} \sum_{x \in \mathcal{D}_{i}} \sum_{c=1}^{C} \bigg\Vert P(X_{i}=c) - P(X_{g}=c) \bigg\Vert,
        \end{aligned}
    \end{equation}
    where $\lambda_{i}$ denotes the Lipschitz constant of PUE $i$ and $\mu$ denotes the upper bound of the gradient of loss.\par
    Let $a_{K-1}^{(m)}$ and $b_{K-1}^{(m)}$ denote
    \begin{equation}
    \label{eq:a}
        \begin{aligned}
            a_{K-1}^{(m)} & = 1 + \frac{\eta}{\lvert \mathcal{P}_{K-1}^{(m)}\rvert} \sum_{i \in \mathcal{P}_{K-1}^{(m)}} \lambda_{i} \\
            & \approx 1 + \frac{\eta}{\lvert \mathcal{P}_{K}^{(m)}\rvert} \sum_{i \in \mathcal{P}_{K}^{(m)}} \lambda_{i} = a^{(m)},
        \end{aligned}
    \end{equation}
    \begin{equation}
    \label{eq:b}
        \begin{aligned}
            & b_{K-1}^{(m)} = \frac{\eta\mu}{\lvert \mathcal{P}_{K-1}^{(m)}\rvert} \sum_{i \in \mathcal{P}_{K-1}^{(m)}} \sum_{x \in \mathcal{D}_{i}} \sum_{c=1}^{C} \lVert P(X_{i}=c) - P(X_{g}=c) \rVert \\
            & \approx \frac{\eta\mu}{\lvert \mathcal{P}_{K}^{(m)}\rvert} \sum_{i \in \mathcal{P}_{K}^{(m)}} \sum_{x \in \mathcal{D}_{i}} \sum_{c=1}^{C} \lVert P(X_{i}=c) - P(X_{g}=c) \rVert = b^{(m)}.
        \end{aligned}
    \end{equation}
    Here, \eqref{eq:a} holds because the upper bound of the gradient is constant according to Assumption \ref{assumption2}, and the Lipschitz constant represents the smallest upper bound of the loss. Moreover, \eqref{eq:b} holds because we assume that the non-IID data of the PUEs follow a symmetric Dirichlet distribution with the same concentration parameters~\citep{ref:noniid2}. Based on \eqref{eq:a} and \eqref{eq:b}, we can finally derive the upper bound of the weight difference as
    \begin{equation}
    \label{pf1:05}
        \begin{aligned}
            & \lVert \mathbf{w}_{t, K}^{(m)} - \mathbf{w}^{(c)}_{t, K} \rVert \\
            & \leq a^{(m)} \lVert \mathbf{w}_{t, K-1}^{(m)} - \mathbf{w}_{t, K-1}^{(c)} \rVert + b^{(m)} \\
            & \leq (a^{(m)})^{2} \lVert \mathbf{w}_{t, K-2}^{(m)} - \mathbf{w}_{t, K-2}^{(c)} \rVert + (1 + a^{(m)})b^{(m)} \\
            & \leq (a^{(m)})^{K} \lVert \mathbf{w}_{t, 0}^{(m)} - \mathbf{w}_{t, 0}^{(c)} \rVert + \sum_{i=1}^{K}(a^{(m)})^{i-1}b^{(m)} \\
            & = (a^{(m)})^{K} \lVert \mathbf{w}_{t, 0}^{(m)} - \mathbf{w}_{t, 0}^{(c)} \rVert \\
            & + \frac{(a^{(m)})^{K} - 1}{a^{(m)} - 1}\frac{\eta\mu}{\lvert \mathcal{P}_{K}^{(m)}\rvert} \sum_{i \in \mathcal{P}_{K}^{(m)}} \sum_{c=1}^{C} \lVert P(X_{i}=c) - P(X_{g}=c) \rVert. \\
        \end{aligned}
    \end{equation}
    Thus, we obtain the following result for \eqref{eq:prop1}.
    \end{proof}
\end{prop}
According to Proposition \ref{prop1}, we can deduce that the upper bound of the weight difference is determined mainly by the weight initialization and probability distance as follows:
\begin{remark}
\label{rem_1}
    \emph{The weight divergence occurs in two major parts. The former and latter terms on the right-hand side of the inequality \eqref{eq:prop1} indicate the parameter initialization and the effect of diffusion, respectively. If the BS initializes the parameter equally in centralized learning and FedDif, the root of the weight divergence only becomes diffusion.}
\end{remark}
\begin{remark}
\label{rem_2}
    \emph{The weight divergence can also be induced by $a^{(m)}$. Excessive diffusion can gradually increase the upper bound of the weight difference, even if diffusion can reduce the probability distance, because $a^{(m)} \geq 1$, $\frac{(a^{(m)})^{K} - 1}{a^{(m)} - 1} \geq 1$.}
\end{remark}
\begin{remark}
\label{rem_3}
    \emph{Diffusion mechanism entails typical issues of existing machine learning, such as overshooting and gradient exploding. Overshooting and gradient exploding by the immoderate learning rate and upper bound of gradients may induce the weight divergence because $\eta\mu/\left\vert \mathcal{P}_{K}^{(m)}\right\vert$ can push up the upper bound of the weight difference. Gradient clipping and learning rate decay can address the overshooting and gradient exploding~\citep{ref:deep_learning}.}
\end{remark}
\begin{remark}
\label{rem_4}
    \emph{The probability distance between the local model and the model of centralized learning in diffusion may induce weight divergence. In other words, approximating the experienced distribution of the local model to the distribution of the IID data can significantly reduce the weight difference.}
\end{remark}
Based on Remark \ref{rem_2}, we can see that selecting a PUE whose private data minimizes the IID distance can lower the upper bound of the weight difference. Inspired by Proposition \ref{prop1} and Remark \ref{rem_3}, we additionally apply a regularization technique that customizes the approach of FedDC~\citep{ref:FedDC}. Different directions of local drifts from those in the previous diffusion round may lead to more severe overfitting, as the non-IID training data in another PUE can distort the optimization trajectory in the opposite direction. We define diffusion drifts as the parameter and gradient differences between the models in the previous and current diffusion rounds as follows:
\begin{equation}
    h_{t, k}^{(m)} = \mathbf{w}_{t, k-1}^{(m)} - \mathbf{w}_{t, k}^{(m)}
\end{equation}
Subsequently, in each local training, we include a regularization term in the loss function using diffusion drifts to reduce the deviation between the local and global models. As the diffusion rounds progress, overfitting may worsen. Therefore, we utilize the reciprocal of the IID distance value for the model's DoL as a coefficient to apply increasing regularization, posing greater regularization as the diffusion rounds proceed. The penalty term based on the customized regularization term can be written as follows:
\begin{equation}
    \begin{aligned}
        & R_{t, k}(\mathbf{w}_{t, k-1}^{(m)}, \mathbf{w}_{t, k}^{(m)}, h_{t, k}^{(m)}) \\
        & = \frac{\mu}{\mathbb{E}\{W_{1}(\psi_{k}^{(m)}, \mathcal{U})\}}\lVert h_{t, k}^{(m)} + \mathbf{w}_{t, k}^{(m)} - \mathbf{w}_{t, k-1}^{(m)} \rVert^{2},
    \end{aligned}
\end{equation}
where 
\begin{equation}
    \begin{aligned}
        & \mathbb{E}\{W_{1}(\psi_{k}^{(m)}, \mathcal{U})\} \\
        & =\frac{1}{\lvert \mathcal{P}_{k}^{(m)}\rvert} \sum_{i \in \mathcal{P}_{k}^{(m)}} \underbrace{\sum_{x \in \mathcal{D}_{i}} \sum_{c=1}^{C} \bigg\Vert P(X_{i}=c) - P(X_{g}=c) \bigg\Vert}_{=W_{1}(\psi_{k}^{(m)}, \mathcal{U})}.
    \end{aligned}
\end{equation}
Consequently, Remark \ref{rem_4} inspires FedDif to reduce the weight difference presented in \eqref{eq:weight_divergence}. \par
Next, we provide a theoretical basis for ensuring that FedDif can reduce the weight difference according to Proposition \ref{prop2}.
\begin{prop}
\label{prop2}
    \emph{
        FedDif can reduce the probability distance between the local model and centralized learning model mentioned in Remark \ref{rem_4} to zero by minimizing the IID distance.
    }
    \begin{proof}
    In the diffusion mechanism, model $m$ should select a PUE whose DSI is the most suitable for minimizing the IID distance of its DoL $\psi_{k}^{(m)}$. That is, each model should enhance the diversity of its DoL. Information entropy can stand for the diversity of DoL, and the diffusion mechanism can be regarded as a mechanism to solve the \emph{``Entropy maximization problem,"} which can maximize the diversity of DoL. Let $H\big(\psi_{k}^{(m)}\big)$ denotes the information entropy of the DoL $\psi_{k}^{(m)}$ of model $m$ at $k$th diffusion round as follows:
    \begin{equation}
        \begin{aligned}
            & H\big(\psi_{k}^{(m)}\big) = -\sum_{c=1}^{C}\psi_{k}^{(m)}[c]\ln\big(\psi_{k}^{(m)}[c]\big) \\
            & = -\sum_{c=1}^{C}\frac{1}{D_{(\mathcal{P}_{k}^{(m)})}}r{\sum_{j \in \mathcal{P}_{k}^{(m)}}}D_{j}\mathbf{d}_{j}[c]\Bigg(\frac{1}{D_{(\mathcal{P}_{k}^{(m)})}}r{\sum_{j \in \mathcal{P}_{k}^{(m)}}}D_{j}\mathbf{d}_{j}[c]\Bigg),
        \end{aligned}
    \end{equation}
    where $\psi_{k}^{(m)}[c]$ and $\mathbf{d}_{j}[c]$ denote the DoL of model $m$ and DSI of PUE $j$ for class $c$, respectively. We state the entropy maximization problem of $H\big(\psi_{k}^{(m)}\big)$ as
    \begin{subequations}
        \begin{align}
            \min_{d_{i_{k}^{(m)}}} \; & -H\big(\psi_{k}^{(m)}\big) \label{eq:lem1Obj} \\
            \textrm{s.t.} \quad & d_{i_{k}^{(m)}}[c] \geq 0, \forall c \label{eq:lem1ConstA} \\
                                & \sum_{c=1}^{C}d_{i_{k}^{(m)}}[c] = 1 \label{eq:lem1ConstB}.
        \end{align}
    \end{subequations}
    Here, we establish a negative entropy minimization problem that is equal to problem \eqref{eq:lem1Obj}, and its optimal solution is provided in Lemma \ref{lemma1}. 
    \begin{lem}
    \label{lemma1}
        \emph{
            The optimal DSI that the model $m$ should select in $k$th diffusion round can be obtained by the principle of maximum entropy as follows:
            \begin{equation}
                d_{i_{k}^{(m)}}^{*}[c] = \frac{1}{D_{i_{k}^{(m)}}}\left(\frac{D_{(\mathcal{P}_{k}^{(m)})}}{C} - D_{(\mathcal{P}_{k-1}^{(m)})}\psi_{k-1}^{(m)*}[c]\right)
            \end{equation}
            where $\sum_{c=1}^{C}d_{i_{k}^{(m)}}^{*}[c]=1$ and $d_{i_{k}^{(m)}}^{*}[c] \geq 0$ hold.
        }
        \begin{proof}
            See \ref{appA}.
        \end{proof}
    \end{lem}
    According to Lemma \ref{lemma1}, the optimal DSI at the $k$th diffusion round is the difference between the number of trained data during the previous diffusion round and the expected number of data if model $m$ trains the IID data during $k$ rounds. That is, FedDif aims to train the model evenly in each class. This can be achieved by minimizing the IID distance, which measures the similarity between the DoL of the model and uniform distribution. However, in the real world, PUEs' dataset is highly heterogeneous; each PUE has a different data size or imbalanced data. Moreover, if the channel condition is poor or there are no neighboring PUEs nearby, the model may not find a user with an optimal DSI. Therefore, we demonstrate that the real-world IID distance can converge to zero when there are sufficient diffusion rounds as in Lemma \ref{lemma2}.
    \begin{lem}
    \label{lemma2}
        \emph{
            The closed form of the IID distance for real-world DoL can be expressed as
            \begin{equation}
            \label{eq:lemma2-1}
                W_{1}(\psi_{k}^{(m)}, \mathcal{U}) = \frac{\left\lVert \phi_{k} - \bar{\phi_{k}} \right\rVert}{D_{\left(\mathcal{P}_{k}^{(m)}\right)}},
            \end{equation}
            where $\phi_{k} =\{\phi_{k}[1], ..., \phi_{k}[C]\} \in \mathbb{R}^{C}$ denotes the variation vector. The average variation is $\bar{\phi_{k}} = \frac{1}{C}\sum_{c=1}^{C}\phi_{k}[c]$. As the number of diffusion rounds increases, the total data size of the subchain increases linearly. In contrast, the variation in the data size of a PUE possessing real-world and optimal DSI at a specific time is an independent and identically distributed value. Therefore, the IID distance for a real-world DoL asymptotically converges to zero as follows:
            \begin{equation}
            \label{eq:lemma2-2}
                \lim_{k \to \infty} W_{1}(\psi_{k}^{m}, \mathcal{U}) = 0,
            \end{equation}
            where $\delta(\psi_{k}^{m}, \mathbf{I})$ is the real-world IID distance.
        }
        \begin{proof}
            See \ref{appB}.
        \end{proof}
    \end{lem}
    We can observe that the IID distance for a real-world DoL converges to zero with sufficient diffusion by Lemma \ref{lemma2}.
    \end{proof}
\end{prop}
\begin{figure*}[ht!]
    \centering
    \includegraphics[width=\textwidth]{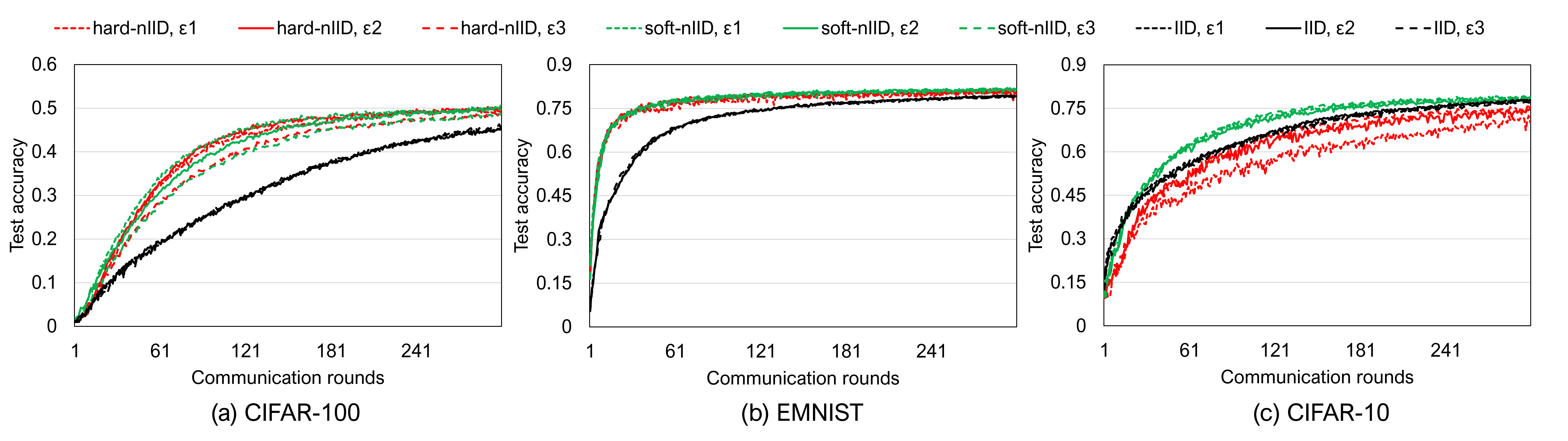}
    \caption{Test accuracy comparison by changing values of concentration parameter and minimum tolerable IID distance. Subfigures (a), (b), and (c) indicate CIFAR-100, EMNIST, and CIFAR-10 datasets, respectively.}
    \label{fig:revised_fig_1}
\end{figure*}
According to Proposition \ref{prop2}, we obtain the convergence point of IID distance for model $m$ after sufficient diffusion iterations. Although FedDif may not discover PUEs with the optimal DSI, it also enables the IID distance to converge to zero by determining the maximum variance in IID distance in the given communication resources. We conclude that FedDif can significantly reduce the weight divergence of FL when there are sufficient diffusion rounds.

\section{Experimental analysis}\label{section6}
In this section, we present the experimental results to evaluate the performance of FedDif. Specifically, we focus on three perspectives: test accuracy, communication costs, and convergence speed. In Section \ref{section6A}, we first introduce the experimental settings and baselines for comparison with FedDif. Then, we investigate the performance of FedDif for different non-IID-ness in Section \ref{section6B}. We compare our methods with the baselines in Section \ref{section6C}. Finally, we discuss the communication efficiency of FedDif compared to the baselines in Section \ref{section6D}.\par

\subsection{Experimental setup}\label{section6A}
\textbf{Benchmark Datasets.}
We consider three benchmark datasets: CIFAR-100, EMNIST, and CIFAR-10 datasets. All of the datasets are split into training and test datasets.
We divide the training dataset into several non-IID data subsets that reflect the two aspects of non-IIDness defined by \citep{ref:FLT}: pathological non-IIDness and quantity skew. Each non-IID data subset is distributed by a Dirichlet distribution, and the non-IID-ness is controlled by the concentration parameter $\alpha$. Note that the dataset approaches the IID when $\alpha$ increases. We set three concentration parameters: 0.3, 0.6, and 100.0, corresponding to hard-nIID, soft-nIID, and IID, respectively. The total quantities per user for all training data subsets and per class of each subset are different.\par

\textbf{Baselines.}
We consider five baseline methods to verify FedDif. First, we compare FedDif with four SOTA methods: FedDC~\citep{ref:FedDC}, FedSwap~\citep{ref:AIoT}, TT-HF~\citep{ref:selected1}, and STC~\citep{ref:selected2}, which address the performance degradation caused by non-IID datasets. FedDC is a regularization method that utilizes parameters and gradient drifts as regularization terms. FedSwap is a model-swapping method that shuffles local models before global aggregation. TT-HF is a user cooperation method that forms several user clusters and cooperatively trains the model between users in the cluster. STC is a model compression method that compresses local models using a sparse ternary compression scheme. Additionally, we compare FedDif with FedAvg~\citep{ref:fedavg}, which is the primary FL method.\par

\textbf{ML tasks.}
To compare the performance of FedDif with SOTA methods, we use a convolutional neural network (CNN) model for image classification tasks with CIFAR-10 and CIFAR-100 datasets. We train a fully connected network (FCN) model for the EMNIST dataset. Although neural networks do not satisfy Assumptions \ref{assumption1} and \ref{assumption2}, we empirically demonstrate that FedDif is still effective for neural networks. We set the learning rate and decay rate to 0.1 and 0.998, respectively, which are also set by \citep{ref:FedDC}. We set the momentum of SGD, batch size, and local training epochs to 0.9, 50, and 1, respectively.\par

\begin{figure*}[ht!]
    \centering
    \includegraphics[width=\textwidth]{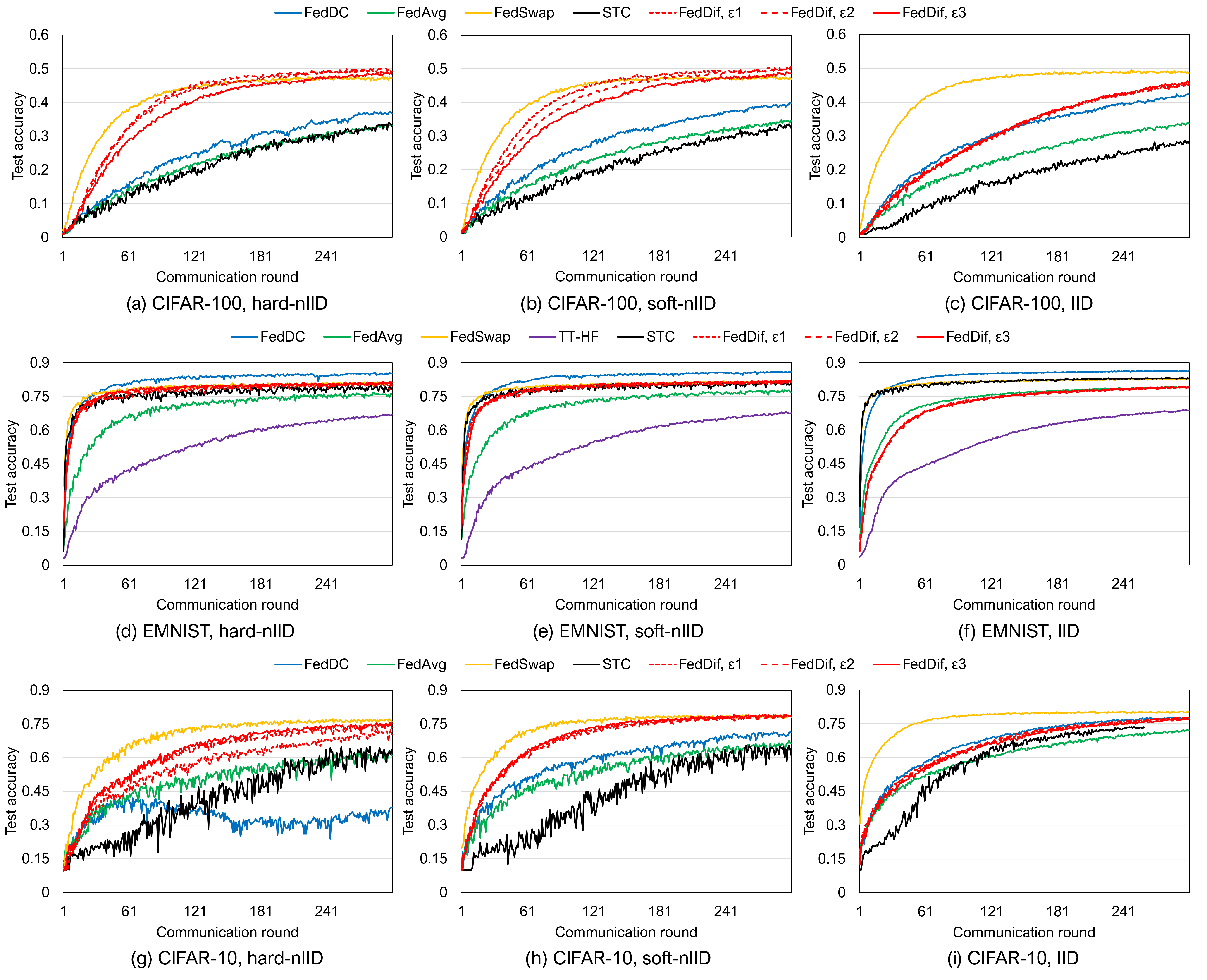}
    \caption{Test accuracy comparison corresponding to the heterogeneity of dataset and ML tasks. Subfigures (a)-(c), (d)-(f), and (g)-(i) indicate CIFAR-100, EMNIST, and CIFAR-10 datasets, respectively.}
    \label{fig:revised_fig_2}
\end{figure*}

\begin{table}[t]
    \caption{Base settings of the minimum tolerable IID distance}
    \begin{tabular}{cccc}
        \toprule
        ML tasks           & $\epsilon1$ & $\epsilon2$ & $\epsilon3$ \\
        \midrule
        CNN with CIFAR-100 & 0.02 & 0.04 & 0.08 \\
        FCN with EMNIST    & 0.04 & 0.08 & 0.12 \\
        CNN with CIFAR-10  & 0.12 & 0.16 & 0.18 \\
        \bottomrule
    \end{tabular}
    \centering
    \label{tab:epsilon_settings}
\end{table}

\subsection{Discussion on the heterogeneity of non-IID data}\label{section6B}
We investigated the performance of FedDif using different heterogeneity of data as shown in Fig. \ref{fig:revised_fig_1}. The heterogeneity of the data is adjusted using different concentration parameters $\alpha$ of the Dirichlet distribution. We set the minimum tolerable QoS $\bar{\gamma}_{\textrm{min}}$ and the minimum tolerable IID distance $\varepsilon$ as 1.0 and 0.04, respectively. We refer to the modulation coding scheme (MCS) index of modulation order two to set the value of $\bar{\gamma}_{\textrm{min}}$~\citep{ref:standard1}.\par
In the case of IID ($\alpha = 100.0$), the models do not require diffusion, as they learn the same data distribution, regardless of which user it is passed to. Therefore, FedDif operates similarly to FedAvg. For hard-nIID and soft-nIID ($\alpha = 0.3, 0.6$), the models require more diffusion rounds to learn the different data distributions. We observed that more models must be transmitted by increasing $\alpha$. Specifically, in the case of IID, because there are no diffusions, FedDif shows a somewhat lower performance owing to insufficient training iterations. We found that the required number of diffusion rounds increases as the heterogeneity of the dataset increases; however, insufficient training iterations can yield lower performance. Rather, the performance may degrade when the models are overtrained.\par
Depending on the complexity of the ML task, the convergence speed can differ owing to the additional diffusion rounds. In FedDif, diffusion is determined by the minimum tolerable IID distance $\epsilon$. The diffusion process gradually reduces the IID distance, and theoretically, it is possible to converge the IID distance to zero. Adversely, as diffusion efficiency decreases, performance improvement relative to the communication costs may become marginal. Therefore, we define the minimum tolerable IID distance to terminate the diffusion rounds once the IID distance is reduced to a certain level. A larger value of $\epsilon$ terminates diffusion earlier. As seen in Table \ref{tab:epsilon_settings}, we set three levels of $\epsilon$ for each ML task, ranging from the case of $\epsilon1$, where models pass through most participants, to the case of $\epsilon2$ where half of the participants are involved, and the case of $\epsilon3$ with minimal participant involvement. By adding diffusion rounds, we observed a visible performance improvement in a highly complex ML task using the CIFAR-100 dataset. However, for the low-complexity ML tasks with EMNIST and CIFAR-10 datasets, it can be found that additional diffusion rounds do not lead to significant performance improvement. Nevertheless, non-IID cases with diffusion improved performance compared with cases without diffusion, such as the IID case. We can see that, in low-complexity ML tasks, diffusion can significantly improve performance, and a small number of diffusion rounds can lead to substantial performance improvements.\par


\begin{sidewaystable*}[t]
    \caption{Comparison of the top-1 test accuracy (\%).\\(300 communication rounds, 100 clients)}
    \begin{tabularx}{0.99\textwidth}{c|ccc|ccc|ccc}
        \toprule
        \multirow{2}{*}{Methods} & \multicolumn{3}{c|}{\multirow{1}{*}{CIFAR-100}} & \multicolumn{3}{c|}{\multirow{1}{*}{EMNIST}} & \multicolumn{3}{c}{\multirow{1}{*}{CIFAR-10}} \\
        \cline{2-10}
        & hard-nIID & soft-nIID & IID & hard-nIID & soft-nIID & IID & hard-nIID & soft-nIID & IID \\
        \midrule\midrule
        \makecell{FedAvg}   & 33.45 & 34.79 & 33.95 & 76.75 & 77.84 & 79.38 & 63.20 & 67.50 & 72.34 \\
        \makecell{FedDC}    & 37.25 & 39.99 & 42.41 & \textbf{85.49} & \textbf{86.00} & \textbf{86.41} & 42.01 & 71.35 & 77.86 \\
        \makecell{FedSwap}  & 47.58 & 47.80 & \textbf{49.56} & 81.63 & 81.91 & 82.98 & \textbf{77.20} & 79.01 & \textbf{80.49} \\
        \makecell{TT-HF}    & N/A & N/A & N/A & 67.00 & 68.06 & 68.94 & N/A & N/A & N/A \\
        \makecell{STC}      & 33.82 & 33.51 & 28.62 & 80.06 & 81.46 & 83.31 & 64.93 & 65.51 & 73.57 \\
        \cline{1-10}
        \makecell{FedDif, $\epsilon1$} & 49.86 & \textbf{50.62} & 45.19 & 80.93 & 81.48 & 79.18 & 72.38 & \textbf{79.19} & 77.83 \\
        \makecell{FedDif, $\epsilon2$} & \textbf{50.25} & 50.24 & 45.36 & 81.71 & 81.80 & 79.49 & 75.31 & 78.90 & 77.89 \\
        \makecell{FedDif, $\epsilon3$} & 48.84 & 48.99 & 46.32 & 81.45 & 82.04 & 79.36 & 75.61 & 79.07 & 77.17 \\
        \bottomrule
    \end{tabularx}
    \centering
    \label{tab:top-1_acc}
\end{sidewaystable*}

\begin{sidewaystable*}[t]
    \caption{Comparison of communication costs.\\(300 communication rounds, CIFAR-100, 100 clients)}
    \small
    \begin{tabularx}{0.985\textwidth}{c|cccc|cccc|cccc}
        \toprule
        \multirow{3}{*}{Methods} & \multicolumn{4}{c|}{\multirow{1}{*}{hard-nIID}} & \multicolumn{4}{c|}{\multirow{1}{*}{soft-nIID}} & \multicolumn{4}{c}{\multirow{1}{*}{IID}} \\
        & \multicolumn{4}{c|}{Target accuracy 33.45\%} & \multicolumn{4}{c|}{Target accuracy 34.79\%} & \multicolumn{4}{c}{Target accuracy 33.95\%} \\
        \cline{2-13}
        & Comm. & Iter. & Cost & Ratio & Comm. & Iter. & Cost & Ratio & Comm. & Iter. & Cost & Ratio \\
        \midrule\midrule
        \makecell{FedAvg}   & 294 & 294 & 27,430 & - & 291 & 291 & 27,150 & - & 299 & 299 & 27,897 & - \\
        \makecell{FedDC}    & 223 & 223 & 20,806 & $\times$1.32 & 204 & 204 & 19,033 & $\times$1.43 & 156 & 156 & 14,555 & $\times$1.92 \\
        \makecell{FedSwap}  & 45 & 450 & 23,092 & $\times$1.19 & 45 & 450 & 20,993 & $\times$1.29 & 38 & 380 & 19,500 & $\times$1.43 \\
        \makecell{STC}      & 295 & 295 & 27,524 & $\times$1.00 & N/A & N/A & N/A & N/A & N/A & N/A & N/A & N/A \\
        \cline{1-13}
        \makecell{FedDif, $\epsilon1$} & 65 & 579 & 30,043 & $\times$0.91 & 62 & 444 & 23,605 & $\times$1.15 & 149 & 149 & 13,902 & $\times$2.01 \\
        \makecell{FedDif, $\epsilon2$} & 62 & 368 & 20,060 & $\times$1.37 & 73 & 362 & 20,293 & $\times$1.34 & 150 & 150 & 13,995 & $\times$1.99 \\
        \makecell{FedDif, $\epsilon3$} & 82 & 328 & \textbf{19,127} & $\times$1.43 & 89 & 267 & \textbf{16,607} & $\times$1.63 & 146 & 146 & \textbf{13,622} & $\times$2.05 \\
        \bottomrule
    \end{tabularx}
    \centering
    \label{tab:comm_cost}
\end{sidewaystable*}

\subsection{Comparison of FedDif with other SOTA methods}\label{section6C}
We validated the performance of FedDif by comparing its test accuracy and communication cost with those of baselines. Note that we only validated TT-HF with the EMNIST dataset because of the limited compatibility of experiments with TT-HF in the latest experimental environment and specified the top-1 accuracy of TT-HF for CIFAR-10 and CIFAR-100 datasets in Table \ref{tab:top-1_acc} as N/A. All experiments were conducted over 300 communication rounds. Fig. \ref{fig:revised_fig_2} depicts the test accuracy for each communication round on the baselines and FedDif. In the case of CIFAR-100, FedSwap exhibited the fastest convergence in terms of the number of communication rounds. Although it appears that FedDif converges slower than FedSwap, the convergence speed of FedDif is higher. This is because FedSwap shuffles local models ten times within a single communication round, as stated by the authors in \citep{ref:AIoT}, and shows more iterations of local training to achieve top-1 accuracy when including the number of shuffling in the total iterations. Detailed comparisons of the convergence rates are discussed later.\par
As shown in Table \ref{tab:top-1_acc}, FedDif achieved the highest top-1 test accuracy in the case of hard-nIID and soft-nIID. Specifically, it achieved 50.25\% and 50.62\% of the top-1 accuracy for hard-nIID and soft-nIID, respectively. Because FedDif does not diffuse models in the case of IID, it may exhibit a relatively lower performance owing to insufficient training iterations. In Table \ref{tab:comm_cost}, for IID, FedSwap required 380 iterations, while FedDif performs approximately 150 iterations. In other words, FedSwap requires roughly 2.53 times the communication cost for a performance improvement of approximately 6.99\%. In the case of CIFAR-100 dataset, given its high task complexity, $\epsilon1$ with higher iterations exhibited the highest performance.\par
For the EMNIST dataset, which had the lowest task complexity, most methods, except FedAvg, converged quickly. Consequently, the potential for performance improvement based on additional iterations is relatively low. For example, FedDC shows a higher top-1 accuracy than FedDif and FedSwap because of the lower task complexity of EMNIST. The CIFAR-10 dataset's task complexity lies between those of EMNIST and CIFAR-100 datasets. In the hard-nIID of CIFAR-10 dataset, FedSwap exhibits approximately 2.1\% higher accuracy than FedDif but requires more iterations, making it less communication efficient. In the case of soft-nIID, FedDif outperformed all baselines. Thus, we conclude that FedDif is adequate for highly complex ML tasks.\par

\subsection{Discussion on the communication efficiency}\label{section6D}
To demonstrate the effectiveness of FedDif, we analyzed its communication efficiency for CIFAR-100 dataset, which has the highest complexity. First, we set the target accuracy
\footnote{Since FedAvg exhibits the lowest performance among baselines, and setting the target accuracy as the peak accuracy of other baselines would make it difficult to compare with methods that perform worse than this target, we chose to set the target accuracy to FedAvg's peak accuracy.}
and define the communication cost as the number of training iterations and bytes transmitted for each method until the target accuracy is achieved. As shown in Table \ref{tab:comm_cost}, \emph{Comm.} and \emph{Iter.} represent the communication rounds and local training iterations required to achieve the target accuracy, respectively. \emph{Cost.} reflects the total bytes of the transmitted models required to attain the target accuracy and \emph{Ratio.} indicates the reduction ratio of the communication cost relative to FedAvg. The unit of the communication cost is megabytes. In the cases of Soft-nIID and IID, STC's communication cost is denoted as N/A because STC did not achieve the target accuracy within the given iteration budget. The communication cost can be calculated as the total number of bytes transmitted and received during the BS, which sends the global model to PUEs and collects local models after local training. For FedDif and FedSwap, we added all the bytes transmitted during model diffusion and shuffling to the communication cost within a communication round.\par
Table \ref{tab:comm_cost} shows that FedSwap requires the fewest communication rounds for all cases of data heterogeneity. This is because FedSwap had the highest number of iterations within a single communication round. During the iterations, both FedSwap and FedDif exhibited higher iterations owing to multiple local model diffusions within a communication round. Consequently, both FedDif and FedSwap require relatively large communication and computing resources from PUEs while significantly reducing the BS's resources related to model broadcasting and aggregation. \footnote{Based on this fact, when AI applications operating on wireless networks are implemented using federated learning, FedDif can significantly conserve the limited radio resources within the network.}\par
From a communication cost perspective, FedDif outperformed all the baselines. Specifically, methods such as FedAvg,  FedDC, and STC, which perform local training only once within a communication round, incur communication costs when the BS sends the model to PUEs and when users transmit the model back to the BS after local training. However, FedDif can minimize the model transmissions between the BS and PUEs, thereby reducing communication costs against other baselines. Compared to FedAvg, in the cases of hard-nIID, soft-nIID, and IID, FedDC, which induces minimum communication cost among baselines, reduced communication costs by 1.32 times, 1.43 times, and 1.92 times, respectively, while FedDif reduced them by 1.43 times, 1.63 times, and 2.05 times, respectively. Whereas FedDC requires the lowest communication cost among the baselines, FedDif reduces the communication cost by 8.78\%, 14.6\%, and 6.85\% for hard-nIID, soft-nIID, and IID, respectively, compared with FedDC.\par

\section{Conclusion and future works}\label{section7}
We proposed a novel diffusion strategy to prevent performance degradation using non-IID data in federated learning. We designed FedDif for the models to obtain knowledge of various local datasets before global aggregation. We regarded every user as non-IID batch data and the models learned different non-IID data by passing them through multiple users before the global aggregation step. Every model can obtain the same effect when trained using a nearly IID dataset. Furthermore, based on the auction theory, we empowered communication efficiency in FedDif to achieve a trade-off between improving learning performance and minimizing communication costs. This study provides several opportunities for future research in this area. For example, FedDif deals with horizontal FL, where the feature set of every dataset is the same. Considering vertical federated learning, in which each user has a different feature set, designing aggregation methods for the feature set on a server will be a crucial topic for future studies.

\section*{Acknowledgement}
This work was supported by Samsung Research Funding \& Incubation Center of Samsung Electronics under Project Number SRFC-TE2103-02 and was supported by Institute for Information \& communications Technology Planning \& Evaluation (IITP) grant funded by the Korea government (MSIT) (No. 2021-0-00368, Development of the 6G Service Targeted AI/ML-based autonomous-Regulating Medium Access Control (6G STAR-MAC)).

\appendix
\section{Implementations of FedDif}

\subsection{Minimum tolerable QoS}
Although the auction can ensure communication-efficient diffusion, the BS may conduct \emph{full diffusion} until the improvement in IID distance converges to zero. Note that full diffusion participates in all PUEs in the diffusion. In full diffusion, PUEs consuming substantial communication resources may exist because distant PUEs can be paired to improve the IID distance. We set the minimum tolerable QoS requirement to constrain the excessive communication resources, expressed as \eqref{eq:ConstE}. However, an isolation problem may occur when no neighboring PUE satisfy the minimum tolerable QoS. The models of isolated PUEs may break the synchronization of the FL or create an undertrained global model. This is similar to the insufficient batch data in centralized learning. Therefore, a higher value of the minimum tolerable QoS may deepen the isolation problem. The D2D communications considering outage probability for PUE pairing is crucial. We pair two PUEs $i$ and $j$ by considering the outage probability as follows:
\begin{equation}
    P_{\textrm{out}}\left(\gamma_{i, j}^{(k)} \leq \gamma_{\textrm{min}}\right) = 1 - \textrm{exp}\Bigg(\frac{-\Big(2^{R_{i, j}^{(k)}} - 1\Big)}{\textrm{SNR}_{i, j}^{(k)}}\Bigg),
\end{equation}
where $R_{i, j}^{(k)} = B_{k}^{(m)}\gamma_{i, j}^{(k)}$ and $\textrm{SNR}_{i, j}^{(k)} = \frac{\vert g_{i, j}^{(k)} \vert^{2} p_{i}^{(k)}}{\sigma^{2}}$ hold \citep{ref:selected1}. In FedDif, the BS only schedules PUEs if and only if their respective outage probabilities satisfy $P_{\textrm{out}}\left(\gamma_{i, j}^{(k)} \leq \gamma_{\textrm{min}}\right) \leq 5\%$ for a given minimum tolerable QoS $\gamma_{\textrm{min}}$. Note that the minimum tolerable QoS is given as a predefined spectral efficiency and can be used to derive the data rate.\par

\subsection{Stop condition}
Another concern is that immoderate diffusion may deteriorate communication efficiency even though a sufficient number of diffusion rounds can reduce the IID distance. In other words, the variance in IID distance may be insufficient compared with the required communication resources in the case of low diffusion efficiency. The minimum tolerable IID distance $\varepsilon$ enables the BS to halt the diffusion and perform the global aggregation step. The halting condition for each model is expressed as $W_{1}\left(\psi_{k}^{(m)}, \mathcal{U}\right) \le \varepsilon$. The number of diffusion rounds should increase when the operator focuses on the learning performance. In other words, the BS can maintain diffusion by employing a low minimum tolerable IID distance. However, the BS should employ a high minimum tolerable IID distance under limited communication resources.\par

\subsection{Optimality of winner selection}
The Kuhn-Munkres algorithm obtains the optimal solution $\mathcal{R}^{*}$ by using the following principle
\begin{equation}
    \label{eq:hungarian_const}    
    \sum_{e \in \mathcal{R}^{*}}c(e) \geq \sum_{e' \in \mathcal{R}'}c(e'), \forall \mathcal{R}' \subseteq \mathcal{M}\times\mathcal{N}_{P}
\end{equation}
where $\mathcal{R}'$ and $\mathcal{R}^{*}$ denote subsets of the feasible and optimal matching sets, respectively. $e$ denotes a single edge; Therefore, the BS can configure the scheduling policy for the diffusion to every PUE by deriving the optimal PUE pairing information $\mathbf{i}_{k}^{*} = \left[\left\{ i : (m, i) \in \mathcal{R}^{*} \right\}\right]$ and required communication resources $\mathbf{B}_{k}^{*} = \left[\left\{ \Tilde{B}^{(m)}_{i, k} : (m, i) \in \mathcal{R}^{*} \right\}\right]$ based on the result of the Kuhn-Munkres algorithm.\par

\subsection{Analysis of the convergence for IID distance and diffusion efficiency}
\begin{figure}[t]
    \centering
    \includegraphics[width=\columnwidth]{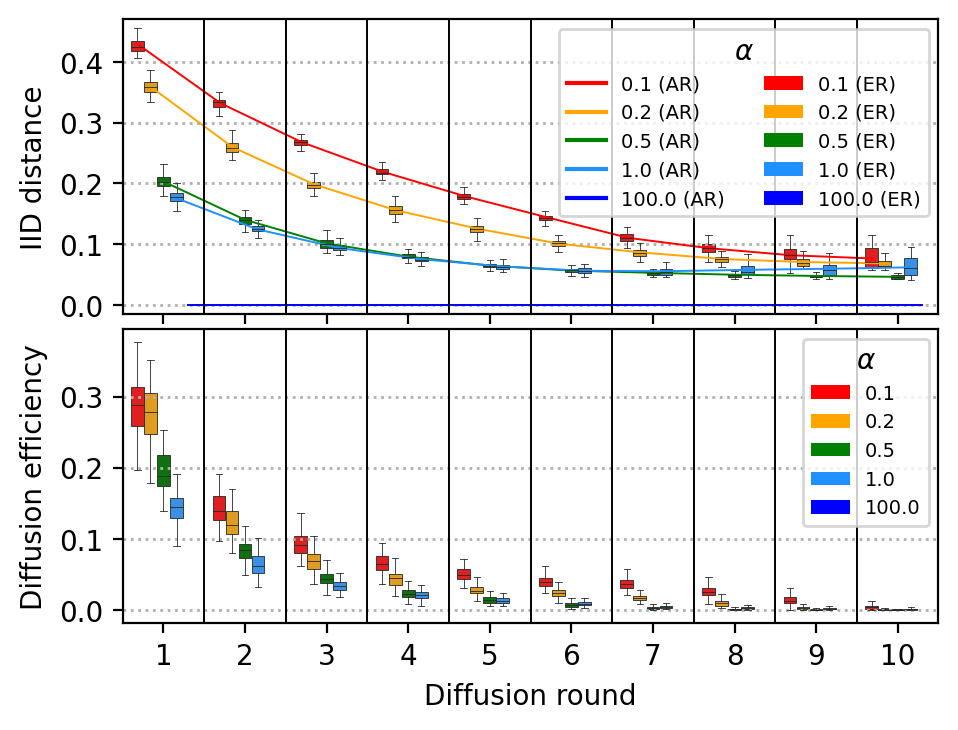}
    \caption{Convergence trends of IID distance and diffusion efficiency by different concentration parameters.}
    \label{fig:exp_1}
\end{figure}

We further investigated the convergence trends of IID distance and diffusion efficiency as the number of diffusion rounds increased, as shown in Fig. \ref{fig:exp_1}. Fig. \ref{fig:exp_1} consists of two results: the analytical results (AR) and experimental results (ER) represented by line and box plots, respectively. The AR of IID distance is computed using the closed form of the IID distance for a real-world DoL in \eqref{eq:lemma2-1}. The AR and ER of IID distance converge to zero as the number of diffusion rounds increases. We divided the universal dataset into ten non-IID datasets by Dirichlet distribution and distribute each non-IID dataset to PUEs. We can see that the IID distances converge in ten diffusion rounds because each model can learn the entire dataset when it learns ten pieces of the universal dataset, namely, ten types of non-IID datasets. In the case of diffusion efficiency, the entire trend was the same as that in the case of IID distance. The diffusion efficiency and IID distance in the case of a high concentration parameter $\alpha$, which means a high degree of non-IID, decrease steeply compared to the low concentration parameter. This implies that FedDif requires more diffusion rounds under a lower concentration parameter, where the dataset is highly biased. Meanwhile, the diffusion mechanism entails the typical issues of existing machine learning, such as overshooting and gradient explosion. Overshooting and gradient explosion by an immoderate learning rate and upper bound of gradients may induce weight divergence because $\eta\mu/\left\vert \mathcal{P}_{K}^{(m)}\right\vert$ can increase the upper bound of the weight difference. Gradient clipping and learning-rate decay can address overshooting and gradient explosions.

\section{Proof of Lemma \ref{lemma1}}\label{appA}
\setcounter{equation}{0}
To obtain the optimal solution to the entropy maximization problem \eqref{eq:lem1Obj}, we first define the Lagrangian function $\mathcal{L}$ on the DSI $\mathbf{d}_{i_{k}^{m}}$, which can be expressed as
\begin{equation}
    \label{eq:appAeq1}
    \begin{aligned}
        & \mathcal{L}(\mathbf{d}_{i_{k}^{(m)}}[c], u) \\
        & = \sum_{c=1}^{C}\frac{1}{D_{\left(\mathcal{P}_{k}^{(m)}\right)}}\sum_{j \in \mathcal{P}_{k}^{(m)}}D_{j}\mathbf{d}_{j}[c]\ln\Bigg(\frac{1}{D_{\left(\mathcal{P}_{k}^{(m)}\right)}}\sum_{j \in \mathcal{P}_{k}^{(m)}}D_{j}\mathbf{d}_{j}[c]\Bigg) \\
        & \quad + u\Bigg(\sum_{c=1}^{C}\mathbf{d}_{i_{k}^{(m)}}[c]-1\Bigg)
    \end{aligned}
\end{equation}
where $u$ denotes the Lagrangian multiplier corresponding to constraint \eqref{eq:lem1ConstA}. The first- and second-order derivatives of the Lagrangian function $\mathcal{L}(\mathbf{d}_{i_{k}^{(m)}}[c], u)$ are given by
\begin{equation}
    \label{eq:appAeq2}
    \begin{aligned}
        & \frac{\partial \mathcal{L}(\mathbf{d}_{i_{k}^{(m)}}[c], u)}{\partial \mathbf{d}_{i_{k}^{(m)}}[c]} \\
        & = \frac{D_{i_{k}^{(m)}}}{D_{(\mathcal{P}_{k}^{(m)})}}\ln\Bigg(\frac{1}{D_{(\mathcal{P}_{k}^{(m)})}}\sum_{j \in \mathcal{P}_{k}^{(m)}}D_{j}\mathbf{d}_{j}[c]\Bigg) + \frac{D_{i_{k}^{(m)}}}{D_{(\mathcal{P}_{k}^{(m)})}} + u
    \end{aligned}
\end{equation}
and
\begin{equation}
    \label{eq:appAeq3}
        \frac{\partial^{2} \mathcal{L}(\mathbf{d}_{i_{k}^{(m)}}[c], u)}{\partial \mathbf{d}_{i_{k}^{(m)}}^{2}[c]} = \frac{D_{i_{k}^{(m)}}^{2}}{\sum_{j \in \mathcal{P}_{k}^{(m)}}D_{j}\mathbf{d}_{j}[c]} \geq 0,
\end{equation}
respectively. From \eqref{eq:appAeq3}, we can see that \eqref{eq:appAeq1} is a convex function. Based on the Karush-Kuhn-Tucker (KKT) conditions, we obtain the optimal solution of the Lagrangian function \eqref{eq:appAeq1} as follows:
\begin{subequations}
    \begin{align}
        \frac{\partial \mathcal{L}(\mathbf{d}_{i_{k}^{(m)}}[c], u)}{\partial \mathbf{d}_{i_{k}^{(m)}}[c]} = 0, \label{eq:KKT1} \\
        u\Bigg(\sum_{c=1}^{C}\mathbf{d}_{i_{k}^{(m)}}[c] - 1\Bigg) = 0, \label{eq:KKT2} \\
        u \geq 0, \label{eq:KKT3}
    \end{align}
\end{subequations}
where \eqref{eq:KKT1}, \eqref{eq:KKT2}, and \eqref{eq:KKT3} denote the \emph{stationary}, \emph{complementary slackness}, and \emph{dual-feasibility} conditions, respectively. By substituting \eqref{eq:appAeq2} into \eqref{eq:KKT1}, we obtain the optimal DSI $\mathbf{d}_{i_{k}^{(m)}}^{*}$ as
\begin{equation}
    \label{eq:appAeq4}
    \frac{D_{i_{k}^{(m)}}}{D_{\left(\mathcal{P}_{k}^{(m)}\right)}}\ln\Bigg(\frac{1}{D_{(\mathcal{P}_{k}^{(m)})}}\sum_{j \in \mathcal{P}_{k}^{(m)}}D_{j}\mathbf{d}_{j}^{*}[c]\Bigg) + \frac{D_{i_{k}^{(m)}}}{D_{\left(\mathcal{P}_{k}^{(m)}\right)}} + u = 0
\end{equation}
Rearranging \eqref{eq:appAeq4}, we obtain
\begin{equation}
    \label{eq:appAeq5}
    \sum_{j \in \mathcal{P}_{k}^{(m)}}D_{j}\mathbf{d}_{j}^{*}[c] = e^{-\left(\frac{D_{\left(\mathcal{P}_{k}^{(m)}\right)}}{D_{i_{k}^{(m)}}}u - \ln D_{\left(\mathcal{P}_{k}^{(m)}\right)} + 1\right)}.
\end{equation}
From the definition of the data size of the diffusion subchain $D_{(\mathcal{P}^{(m)}_{K_{t}})}=\sum_{i \in \mathcal{P}^{(m)}_{K_{t}}}D_{i}$, we can separate the left side of the equation \eqref{eq:appAeq5} as
\begin{equation}
    \label{eq:appAeq6}
    \sum_{j \in \mathcal{P}_{k}^{(m)}}D_{j}\mathbf{d}_{j}[c] = \sum_{j \in \mathcal{P}_{k-1}^{(m)}}D_{j}\mathbf{d}_{j}[c] + D_{i_{k}^{(m)}}\mathbf{d}_{i_{k}^{(m)}}[c]
\end{equation}
Based on the equation \eqref{eq:appAeq6}, we can rewrite \eqref{eq:appAeq5} as follows:
\begin{equation}
    \label{eq:appAeq7}
    \begin{aligned}
        & \mathbf{d}_{i_{k}^{(m)}}^{*}[c] \\
        & = \frac{1}{D_{i_{k}^{(m)}}}\Bigg(e^{-\left(\frac{D_{\left(\mathcal{P}_{k}^{(m)}\right)}}{D_{i_{k}^{(m)}}}u - \ln D_{\left(\mathcal{P}_{k}^{(m)}\right)} + 1\right)} - \sum_{j \in \mathcal{P}_{k-1}^{(m)}}D_{j}\mathbf{d}_{j}^{*}[c]\Bigg).
    \end{aligned}
\end{equation}
Here, we determine the optimal Lagrangian multiplier $u^{*}$ using the complementary slackness condition in \eqref{eq:KKT2} as follows:
\begin{equation}
    \label{eq:appAeq8}
    \sum_{c=1}^{C}\frac{1}{D_{i_{k}^{(m)}}}\Bigg(e^{-\left(\frac{D_{\left(\mathcal{P}_{k}^{(m)}\right)}}{D_{i_{k}^{(m)}}}u^{*} - \ln D_{\left(\mathcal{P}_{k}^{(m)}\right)} + 1\right)} - \sum_{j \in \mathcal{P}_{k-1}^{(m)}}D_{j}\mathbf{d}_{j}^{*}[c]\Bigg) = 1.
\end{equation}
Based on the definition of the data size of the diffusion subchain, we can rearrange \eqref{eq:appAeq8} as
\begin{equation}
    \label{eq:appAeq9}
    \begin{aligned}
        Ce^{-\left(\frac{D_{\left(\mathcal{P}_{k}^{(m)}\right)}}{D_{i_{k}^{(m)}}}u^{*} - \ln D_{\left(\mathcal{P}_{k}^{(m)}\right)} + 1\right)} & = D_{\left(\mathcal{P}_{k-1}^{(m)}\right)} + D_{i_{k}^{(m)}} \\
        & = D_{\left(\mathcal{P}_{k}^{(m)}\right)},
    \end{aligned}
\end{equation}
where the total data size of the diffusion subchain $\mathcal{P}_{k-1}^{(m)}$ is defined by $D_{\left(\mathcal{P}_{k-1}^{(m)}\right)} = \sum_{c=1}^{C}\sum_{j \in \mathcal{P}_{k-1}^{(m)}}D_{j}\mathbf{d}_{j}[c]$. From \eqref{eq:appAeq9}, we can determine the Lagrangian multiplier $u^{*}$ as
\begin{equation}
    \label{eq:appAeq10}
    u^{*} = \frac{D_{i_{k}^{(m)}}}{D_{\left(\mathcal{P}_{k}^{(m)}\right)}}\left( \ln C - 1 \right)
\end{equation}
By substituting \eqref{eq:appAeq10} into \eqref{eq:appAeq7}, we finally obtain the optimal DSI of model $m$ in the $k$th diffusion round as
\begin{equation}
    \label{eq:appAeq11}
    \mathbf{d}_{i_{k}^{(m)}}^{*}[c] = \frac{1}{D_{i_{k}^{(m)}}}\left(\frac{D_{\left(\mathcal{P}_{k}^{(m)}\right)}}{C} - D_{\left(\mathcal{P}_{k-1}^{(m)}\right)}\psi_{k-1}^{(m)}[c]\right).
\end{equation}
where the relationship $D_{\left(\mathcal{P}_{k-1}^{(m)}\right)}\psi_{k-1}^{(m)}[c] = \sum_{j \in \mathcal{P}_{k-1}^{(m)}}D_{j}\mathbf{d}_{j}[c]$ holds.\par
Furthermore, based on the constraint \eqref{eq:lem1ConstA}, we can obtain the lower bound of the feasible size of the dataset for PUE $i_{k}^{(m)}$ which has an optimal DSI. Suppose that the data size of PUE $i_{k}^{(m)}$ satisfies $D_{i_{k}^{(m)}} \geq 0$. Then, we have the following relationship based on constraint \eqref{eq:lem1ConstA}.
\begin{equation}
\label{eq:appAeq12}
    \frac{D_{\left(\mathcal{P}_{k}^{(m)}\right)}}{C} - D_{\left(\mathcal{P}_{k-1}^{(m)}\right)}\psi_{k-1}^{(m)*}[c] \geq 0.
\end{equation}
By rearranging \eqref{eq:appAeq12} in terms of data size $D_{i_{k}^{(m)}}$, we obtain
\begin{equation}
\label{eq:appAeq13}
    \frac{D_{\left(\mathcal{P}_{k-1}^{(m)}\right)} + D_{i_{k}^{(m)}}}{C} - D_{\left(\mathcal{P}_{k-1}^{(m)}\right)}\psi_{k-1}^{(m)*}[c] \geq 0
\end{equation}
\begin{equation}
\label{eq:appAeq14}
    D_{i_{k}^{(m)}} \geq CD_{\mathcal{P}_{k-1}^{(m)}}\psi_{k-1}^{(m)}[c] - D_{\mathcal{P}_{k-1}^{(m)}}
\end{equation}
From inequality \eqref{eq:appAeq14}, we can see that because $D_{i_{k}^{(m)}}$ is always positive for the DoL of every class, $D_{i_{k}^{(m)}}$ should be greater than the boundary value derived as the maximum value among DoL for each class. Thus, we can obtain the lower bound of the feasible data size for each PUE by using Corollary \ref{cor1}.
\begin{cor}
\label{cor1}
    \emph{
        The lower bound of the feasible dataset size for PUE $i_{k}^{(m)}$ with the optimal DSI is
        \begin{equation}
            D_{i_{k}^{(m)}} \geq \max_{c}\left\{CD_{\mathcal{P}_{k-1}^{(m)}}\psi_{k-1}^{(m)}[c] - D_{\mathcal{P}_{k-1}^{(m)}}\right\},
        \end{equation}
        where $C$ denotes the number of classes.
    }
\end{cor}

\section{Proof of Lemma \ref{lemma2}}\label{appB}
\setcounter{equation}{0}
We find where the IID distance for the real-world DoL $\psi_{k}^{(m)}$ converges as there are sufficient diffusion rounds. First, the IID distance for the real-world DoL $\psi_{k}^{(m)}$ is denoted as follows:
\begin{equation}
    \label{eq:appBeq1}
    \begin{aligned}
        W_{1}(\psi_{k}^{(m)}, \mathcal{U}) & = \left\lVert \psi_{k}^{(m)} - \frac{1}{C}\mathbf{1} \right\rVert \\
        & = \sqrt{\sum_{c=1}^{C}\bigg(\psi_{k}^{(m)}[c] - \frac{1}{C}\bigg)^{2}}
    \end{aligned}
\end{equation}
Two cases exist in the $k$th diffusion round. In one case, model $m$ discovers the PUE with the optimal DSI. In another case, model $m$ does not discover the PUE with the optimal DSI. In other words, the model discovers the PUE using real-world DSI. Suppose that there exists a difference, namely, a variation $\phi_{k}[c]$ for two data sizes between the PUE possessing the real-world DSI and the optimal DSI as follows:
\begin{equation}
    \label{eq:appBeq2}
    \sum_{c=1}^{C}\phi_{k}[c] = D_{i_{k}^{(m)}} - D_{i_{k}^{(m)}}^{*} = D_{\left(\mathcal{P}_{k}^{(m)}\right)} - D_{\left(\mathcal{P}_{k}^{(m)}\right)}^{*},
\end{equation}
where $D_{i_{k}^{(m)}}$ and $D_{i_{k}^{(m)}}^{*}$ denote the data sizes of the PUE with real-world DSI and optimal DSI, respectively. $D_{i_{k}^{(m)}}$ and $D_{i_{k}^{(m)}}^{*}$ denote the total data size of the subchains, including the PUE with real-world and optimal DSI, respectively.  Note that we focus on how the IID distance is derived according to the PUE participating in the $k$th diffusion round, regardless of the data trained by the model until the $(k-1)$th diffusion round. In other words, the total data size of the sub-chain in the $(k-1)$th diffusion round $D_{i_{k-1}^{(m)}}$ is the same. Thus, we can formulate a real-world DSI as follows:
\begin{equation}
    \label{eq:appBeq3}
    \mathbf{d}_{i_{k}^{(m)}}[c] = \frac{1}{D_{i_{k}^{(m)}}}\Bigg(\frac{D_{\left(\mathcal{P}_{k}^{(m)}\right)}^{*}}{C} - D_{\left(\mathcal{P}_{k-1}^{(m)}\right)}\psi_{k-1}^{(m)}[c] + \phi_{k}[c]\Bigg),
\end{equation}
where the relationship $D_{\left(\mathcal{P}_{k}^{(m)}\right)}^{*} = D_{\left(\mathcal{P}_{k-1}^{(m)}\right)} + D_{i_{k}^{(m)}}^{*}$ holds. We then obtain the DoL of model $m$ when the model trains the real-world dataset in the $k$th diffusion round as
\begin{equation}
    \label{eq:appBeq4}
        \psi_{k}^{(m)}[c] = \frac{1}{D_{\left(\mathcal{P}_{k}^{(m)}\right)}}\Bigg(D_{\left(\mathcal{P}_{k-1}^{(m)}\right)}\psi_{k-1}^{(m)}[c] + D_{i_{k}^{(m)}}\mathbf{d}_{i_{k}^{(m)}}[c]\Bigg).
\end{equation}
By substituting \eqref{eq:appBeq3} into \eqref{eq:appBeq4}, we obtain
\begin{equation}
    \label{eq:appBeq5}
     \psi_{k}^{(m)}[c] = \frac{1}{D_{\left(\mathcal{P}_{k}^{(m)}\right)}}\Bigg(\frac{D_{\left(\mathcal{P}_{k}^{(m)}\right)}^{*}}{C} + \phi_{k}[c]\Bigg).
\end{equation}
Based on \eqref{eq:appBeq2} and plugging \eqref{eq:appBeq5} into \eqref{eq:appBeq1}, we derive the closed form of the IID distance for a real-world DoL at the $k$th diffusion round as follows:
\begin{equation}
    \label{eq:appBeq6}
    \begin{aligned}
        W_{1}&(\psi_{k}^{(m)}, \mathcal{U}) = \sqrt{\sum_{c=1}^{C}\bigg(\psi_{k}^{(m)}[c] - \frac{1}{C}\bigg)^{2}} \\
        & = \sqrt{\sum_{c=1}^{C}\Bigg(\frac{1}{D_{\left(\mathcal{P}_{k}^{(m)}\right)}}\Bigg(\frac{D_{\left(\mathcal{P}_{k}^{(m)}\right)}^{*}}{C} + \phi_{k}[c]\Bigg) - \frac{1}{C}\Bigg)^{2}} \\
        & = \sqrt{\sum_{c=1}^{C}\Bigg(\frac{D_{\left(\mathcal{P}_{k}^{(m)}\right)}^{*} - D_{\left(\mathcal{P}_{k}^{(m)}\right)} + C\phi_{k}[c]}{CD_{\left(\mathcal{P}_{k}^{(m)}\right)}}\Bigg)^{2}} \\
        & = \sqrt{\sum_{c=1}^{C}\Bigg(\frac{C\phi_{k}[c]-\sum_{c=1}^{C}\phi_{k}[c]}{CD_{\left(\mathcal{P}_{k}^{(m)}\right)}}\Bigg)^{2}} \\
        & = \frac{1}{D_{\left(\mathcal{P}_{k}^{(m)}\right)}}\sqrt{\sum_{c=1}^{C}\bigg(\phi_{k}[c]-\frac{1}{C}\sum_{c=1}^{C}\phi_{k}[c]\bigg)^{2}} \\
        & = \frac{1}{D_{\left(\mathcal{P}_{k}^{(m)}\right)}}\left\lVert \phi_{k} - \bar{\phi_{k}} \right\rVert,
    \end{aligned}
\end{equation}
where $\phi_{k} =\{\phi_{k}[1], ..., \phi_{k}[C]\} \in \mathbb{R}^{C}$ denotes the variation vector. The average variation is denoted as $\bar{\phi_{k}} = \frac{1}{C}\sum_{c=1}^{C}\phi_{k}[c]$. The total data size of the diffusion chain $D_{\left(\mathcal{P}_{k}^{(m)}\right)}$ increased linearly with the number of diffusion rounds. However, the variation in $\left\lVert \phi_{k} - \bar{\phi_{k}} \right\rVert$ is constant with i.i.d. for different diffusion rounds. Therefore, we can see that the IID distance for the real-world DoL $\psi_{k}^{(m)}$ asymptotically converges to zero as follows:
\begin{equation}
    \label{eq:appBeq8}
    \lim_{k \to \infty}W_{1}(\psi_{k}^{(m)}, \mathcal{U}) = \lim_{k \to \infty}\frac{\left\lVert \phi_{k} - \bar{\phi_{k}} \right\rVert}{D_{\left(\mathcal{P}_{k}^{(m)}\right)}} = 0.
\end{equation}
\section{Additional investigation of the gap between the simulation and analytical results}\label{appC-G}
\begin{figure}[t]
    \centering
    \includegraphics[width=\columnwidth]{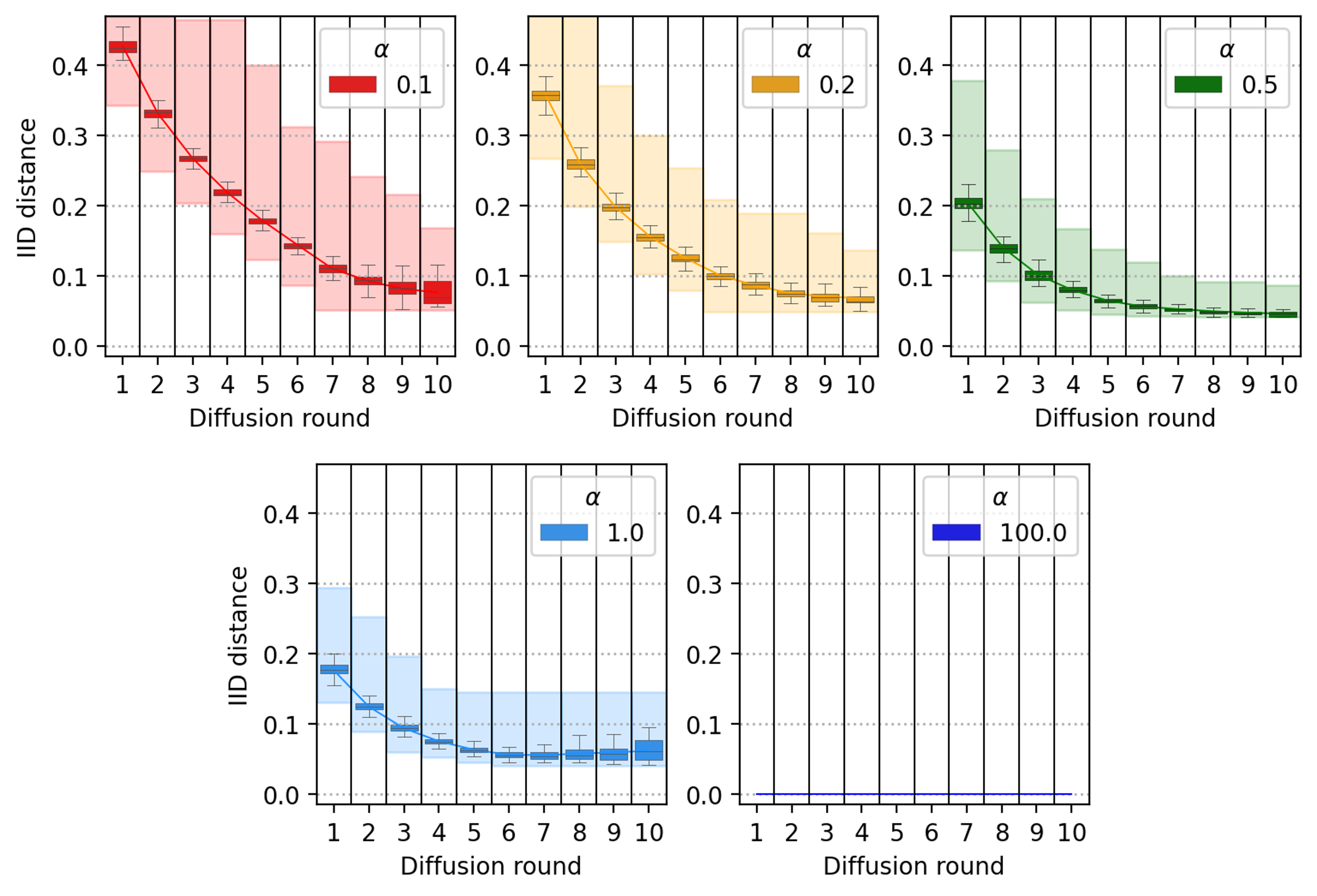}
    \caption{Convergence trends of IID distance by different concentration parameters $\alpha$. Settings: $\alpha$ = 1.0, $\varepsilon$ = 0.04, $\bar{\gamma}_{\textrm{min}}$ = 1.0.}
    \label{fig:add_exp_15}
\end{figure}
Because users have more heavily imbalanced data, the IID distance decreases steeply, starting from a higher point. This is because the more biased the model, the faster it becomes affected by new data that have not been seen before. Fig. \ref{fig:add_exp_15} describes the convergence trends of IID distance by different concentration parameters. Note that the case $\alpha=100.0$ is the same as the distribution of the IID dataset. Each box plot and the shaded area around the box plots represents the experimental and analytical boundaries of IID distance, respectively. We confirm that Lemma 2 holds and that the models do not need to diffuse in a network where every user has the IID dataset.

\bibliographystyle{elsarticle-harv} 
\bibliography{cas-refs}





\end{document}